\documentclass[aps,prx,twocolumn,showpacs,superscriptaddress]{revtex4-1}
\usepackage[english]{babel} 
\usepackage{amssymb}
\usepackage{amsmath}
\usepackage{txfonts}
\usepackage{mathdots}
\usepackage[classicReIm]{kpfonts}
\usepackage[dvips]{graphicx}
\usepackage{epsfig}
\usepackage{graphicx}
\usepackage{array}
\usepackage{amssymb}
\usepackage{amsfonts}
\usepackage{amsmath}
\usepackage{mathrsfs}
\usepackage{color}
\usepackage{booktabs}
\usepackage{threeparttable}
\usepackage{multirow}
\usepackage{subfigure}
\usepackage{epsfig}
\usepackage{threeparttable}
\usepackage{chngpage}
\usepackage{float}
\usepackage{color}
\usepackage{bm}
\usepackage{graphicx}
\usepackage[toc]{appendix}

\newcommand{\beginsupplement}{%
        \setcounter{table}{0}
        \renewcommand{\thetable}{S\arabic{table}}%
        \setcounter{figure}{0}
        \renewcommand{\thefigure}{S\arabic{figure}}%
     }

\begin{document}

\title{Extended range of dipole-dipole interactions in periodically structured photonic media}

\author{Lei Ying}
\affiliation{Department of Electrical and Computer Engineering, University of Wisconsin, Madison, Wisconsin 53706, USA}

\author{Ming Zhou}
\affiliation{Department of Electrical and Computer Engineering, University of Wisconsin, Madison, Wisconsin 53706, USA}

\author{Michael Mattei}
\affiliation{Department of Chemistry, University of Wisconsin, Madison, Wisconsin 53706, USA}

\author{Boyuan Liu}
\affiliation{Department of Electrical and Computer Engineering, University of Wisconsin, Madison, Wisconsin 53706, USA}

\author{Paul Campagnola}
\affiliation{Department of Biomedical Engineering, University of Wisconsin, Madison, Wisconsin 53706, USA}

\author{Randall H. Goldsmith}
\affiliation{Department of Chemistry, University of Wisconsin, Madison, Wisconsin 53706, USA}

\author{Zongfu Yu} \email{zyu54@wisc.edu}
\affiliation{Department of Electrical and Computer Engineering, University of Wisconsin, Madison, Wisconsin 53706, USA}

\date{\today}

\begin{abstract}
The interaction between quantum two-level systems is typically short-range in free space and most photonic environments. Here we show that diminishing momentum isosurfaces with equal frequencies can create a significantly extended range of interaction between distant quantum systems. The extended range is robust and does not rely on a specific location or orientation of the transition dipoles. A general relation between the interaction range and properties of the isosurface is described for structured photonic media. It provides a new way to mediate long-range quantum behavior.
\end{abstract}

\maketitle

\begin{figure}
\begin{center}
\epsfig{figure=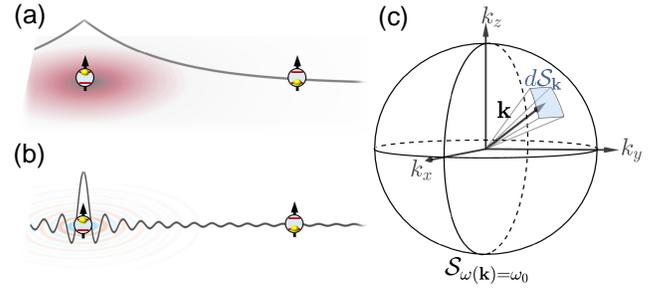,width=\linewidth}
\end{center}
\caption{Schematics of interactions between two TLSs mediated by (a) evanescent near-field modes, (b) propagating far-field modes. (c) Momentum isosurface $S_{\omega \left(\mathbf{k}\right)=\omega_0}$ with equal frequencies $\omega_0$ and $dS_\mathbf{k}$ is a small surface element. }
\label{fig:schematic}
\end{figure}

The resonant dipole-dipole interaction between two quantum two-level systems (TLS) is typically short-range. There has been strong interest in realizing long-range interactions to exploit collective physics such as
superradiance~\cite{scully2009super,solano2017super},
collective frequency shift~\cite{meir2014cooperative},
F$\ddot{\mathrm{o}}$rster resonance energy transfer~\cite{clegg1995fluorescence,garcia2017long},
and quantum entanglement~\cite{van2013photon,burkard2006ultra,petrosyan2008quantum,mingaleev2000long,gonzalez2011entanglement,shahmoon2013nonradiative,hood2016atom}. The ability to modulate the distance dependence of these processes could have potential applications in quantum information processing~\cite{imamog1999quantum,petrosyan2008quantum} and energy conversion~\cite{maxwell2013storage}. Two components contribute to the interaction: the evanescent near fields and the propagating far fields (Fig.~\ref{fig:schematic}a\&b). To enable long-range interaction from the evanescent fields, one could use evanescent fields with a long tail, such as defect modes in the photonic bandgap~\cite{kurizki1990two, douglas2015quantum, notararigo2018resonance}. However, it is less obvious how to engineer propagating far fields to enable long-range interaction. It is the goal of this letter to provide a new perspective to understand the general physical mechanism that is responsible for long-range interaction induced by propagating far fields, and identify photonic structures that are capable of extending the interaction range.

In free space, the range of far-field interaction is limited to the wavelength scale. When the wavelength is long, such as in index-near-zero materials~\cite{fleury2013enhanced,mahmoud2017dipole,liberal2018multiqubit,gundogdu2015asymmetric,serebryannikov2019embedded}, the interaction range can increase proportionally. However, there are a few intriguing examples where the interaction range extends beyond the effective wavelength. These include low-dimensional spaces, such as photonic crystal waveguides and fibers~\cite{sato2012strong, le2017nanofiber,solano2017super,lecamp2007very,hughes2007coupled,yao2009macroscopic,minkov2013radiative,hung2013trapped,vasco2014long}, or hyperbolic materials in selected directions~\cite{biehs2016long,cortes2017super}. These interesting but isolated examples heavily rely on very specific configurations. Thus, it is difficult to generalize the theoretical treatments to identify the underlying physics, which unfortunately remains elusive.
In this letter, we show the deep connection between the interaction range and the size and shape of the isofrequency surface in momentum space. It can be generalized to a broad range of physical systems and can reveal new systems capable of realizing long-range interactions.

We begin by examining the interaction between two TLSs over a long distance. The TLSs are embedded in a photonic environment that can be described by a dispersion relation $\omega =\omega\left( \mathbf{k} \right)$. For example, in free space, $\omega = c|\mathbf{k}|=ck$, where $c$ is the speed of light. Other dispersion relations can be seen in metamaterials, photonic crystals or waveguides. In general, the Hamiltonian of the TLSs and the photonic modes is given by~\cite{bay1997atom}
\begin{equation}
\begin{split}
\mathcal{H}= &
\sum_{i=\mathrm{1,2}} {\omega}_0{{\hat{\sigma }}^\dag_i\hat{\sigma}}_i
+\sum_{\boldsymbol{\mathrm{k}}}{{\omega }_{\boldsymbol{\mathrm{k}}}{\hat{a}}^\dag_{\boldsymbol{\mathrm{k}} }{\hat{a}}_{\boldsymbol{\mathrm{k}} }} \\
+ &
i\sum_{i\mathrm{=1,2}}\sum_\mathbf{k}
\Big[ i g_{\mathbf{k}}(\mathbf{r}_i)
   \left(\hat{\sigma}_{i}^{\dag} + \hat{\sigma}_{i}\right) \hat{a}_{\mathbf{k}}e^{i\mathbf{k}\cdot \mathbf{r}_i} + \mathrm{H.c.} \Big],
\end{split}
\end{equation}
where $\omega_0$ is the resonant transition frequency of TLSs. $\hat{\sigma}^\dag_i(\hat{\sigma}_i)$ is the raising (lowering) operator of $i$th TLS. $\omega _\mathbf{k}$ and $\hat{a}^\dag_{\mathbf{k}}\left({\hat{a}}_{\mathbf{k}}\right)$ are the frequency and creation (annihilation) operator of photons, respectively. $g_{\mathbf{k}}\left(\mathbf{r}_i\right)=\sqrt{\omega_k/2{\varepsilon}_0V}{\boldsymbol{\mathrm{\muup }}}_i\cdot {\boldsymbol{\mathrm{\epsilonup}}}_{\mathbf{k}}$ is the coupling between the $i$th TLS and the photonic mode $\mathbf{k}$, where ${\boldsymbol{\mathrm{\muup}}}_i$ is the transition dipole moment of the $i$th TLS and ${\boldsymbol{\mathrm{\epsilonup}}}_{\boldsymbol{\mathrm{k}}}$ is the polarization direction of the photonic mode $\mathbf{k}$ . One can derive the radiative interaction $\Gamma = \Gamma_{\mathrm{Re}}+i\Gamma_{\mathrm{Im}}$ between two TLSs based on the above Hamiltonian. The real and imaginary parts describe the cooperative decay rate and cooperative energy shift, respectively. The focus of this letter will be the cooperative decay rate. Similar conclusions can be drawn for the cooperative energy shift.

We first provide a graphic illustration of why the interaction between TLSs is short-range in free space. Unlike most theoretical treatments used in the literature~\cite{cortes2017super}, we do not use the Green's function method to describe the radiative environment. Instead, we try to keep all radiative modes in their explicit forms in order to gain a more intuitive picture. As shown in Section I of Supplementary Material (SM), the real part of the radiative interaction between TLSs can be expressed in the following form:
\begin{equation}
\mathrm{\Gamma}_{\mathrm{Re}}
=\iint_{S_{{\omega }_0\left(\boldsymbol{\mathrm{k}}\right)}}{{{\rho }_{\boldsymbol{k}}e}^{i\boldsymbol{\mathrm{k}}\cdot \boldsymbol{\mathrm{R}}}dS_{\boldsymbol{\mathrm{k}}}}.
\end{equation}
The integral is performed on an isosurface in momentum space, i.e. all wavevectors $\mathbf{k}$ that satisfy $\omega\left(\boldsymbol{\mathrm{k}}\right)=\omega_0$. The integrand includes two terms. The first term is simply a polarization factor $\rho_{\mathbf{k}}=\frac{\omega _0}{16\varepsilon_0\pi^2v_g \left( \mathbf{k} \right) }\left({\boldsymbol{\mathrm{\muup}}}_1\cdot {\boldsymbol{\mathrm{\epsilonup }}}_{\boldsymbol{\mathrm{k}}}\right)^\ast
{\left({\boldsymbol{\mathrm{\muup }}}_2\cdot {\boldsymbol{\mathrm{\epsilonup}}}_{\mathbf{k}}\right)}$, which describes the relative orientation of the transition dipole $\boldsymbol{\mathrm{\muup}}$ and the polarization of the electric field $\boldsymbol{\mathrm{\epsilonup}}$. Here $v_g(\mathbf{k})$ is the group velocity of mode $\mathbf{k}$.
For degenerate polarization states, the integration should also include all polarizations.
Since the polarization factor ${\rho}_{\mathbf{k}}$ is independent of the inter-TLS distance, it does not affect the interaction range. It is the second term, $e^{i \mathbf{k}\cdot \mathbf{R}}$, that plays the critical role in the physics of the interaction range. Here $\mathbf{R}=\mathbf{r}_1-\mathbf{r}_2$ is the distance vector between the two TLSs. The integrand $\rho_\mathbf{k} e^{i\mathbf{k}\cdot \mathbf{R}}$ is a fast oscillating function, which generally results in cancellation of the integration when the inter-TLS distance $R$ is large. Therefore, the interaction is always short-range. We can see this effect in Fig. 2a. Here we consider two TLSs in free space. The spherical isosurface has a radius of $k=|\mathbf{k}|={\omega}_0/c$. The real part of $\rho_{k} e^{i\mathbf{k}\cdot \mathbf{R}}$ is plotted on the isosurface. When $R=10\lambda$, there are rapid oscillations as $\mathbf{k}$ varies on the isosurface. The resulting value of the integral is small, and therefore the interaction is weak at this long distance. When the inter-TLS distance is small, for example $R=0.3\lambda$, the oscillation is slow (Figure 2c), leading to a sizeable value of the integral and thus a strong interaction. The interaction decays as the distance $R$ grows (Fig. 2d).

\begin{figure}
\begin{center}
\epsfig{figure=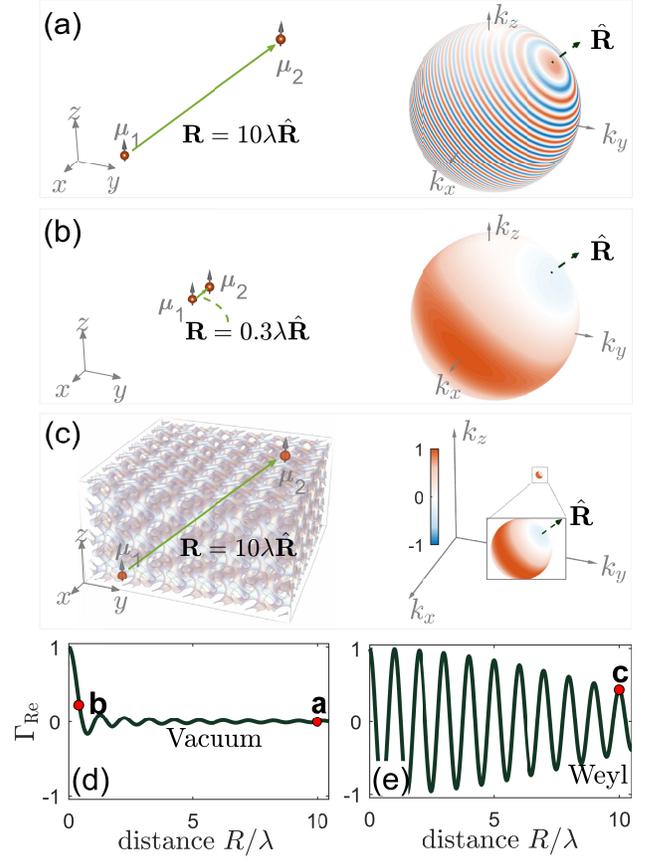,width=\linewidth }
\end{center}
\caption{
(a) Two dipolar quantum transitions spaced by a distance $R=10\lambda$ in free space, where $\lambda=2\pi c/\omega$. The right panel shows the isosurface for the transition frequency in momentum space. The real part of the integrand $\rho_\mathbf{k} e^{i\mathbf{k}\cdot \mathbf{R} }$ is plotted on the isosurface. Red and blue colors indicate positive and negative maximum,respectively. A long $R$ leads to fast oscillation and cancellation of the integral over the isosurface. (b) Similar to (a) but with a shorter distance $R= 0.3\lambda$ and thus slow oscillation on the isosurface. (c) The situation can change significantly if two quantum transitions are placed in a general photonic environment, such as Weyl photonic crystal, where the isosurface can be very small. Here $R = 10\lambda$. The isosurface has a radius of $q=|\mathbf{k}-\mathbf{k}_c|$. The inset in the right panel shows the zoom-in view of the small isosurface, showing that even a large $R$ may not result significant cancellation due to small isosurface size. $\hat{\mathbf{R}}$ in (a-c) is fixed as $\left(1,0,1 \right)/\sqrt{2}$.
(d) \& (e) The real part of radiative interaction, normalized by $\Gamma_\mathrm{Re}(R=0)$, as a function of distance between two TLSs in free space and the Weyl photonic crystal, respectively. Red dots correspond to the cases in (a), (b), and (c), respectively. }
\label{fig:fig2}
\end{figure}

\begin{figure*}
\begin{center}
\epsfig{figure=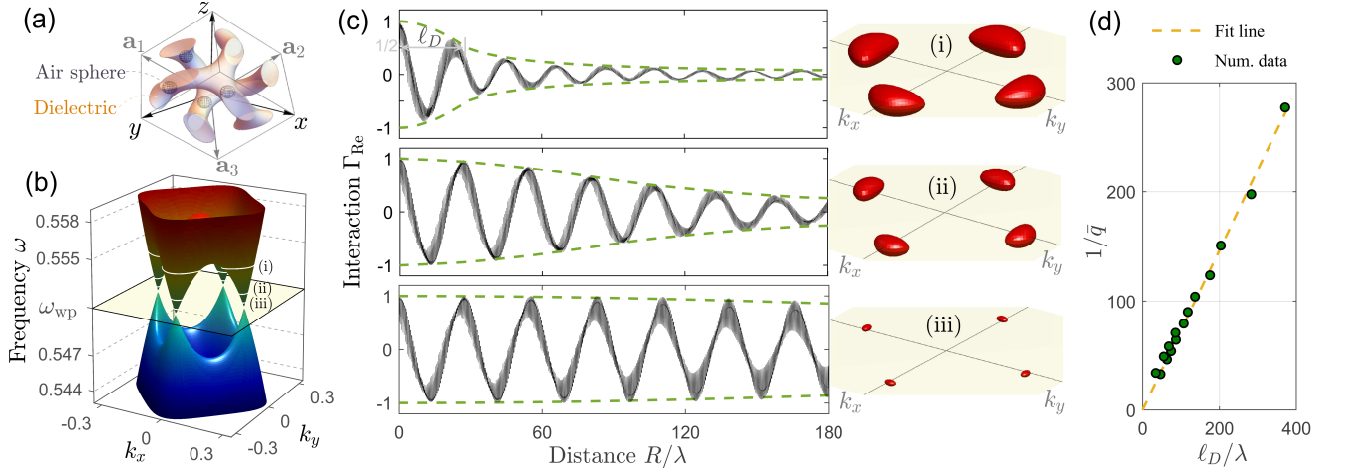,width=\linewidth}
\end{center}
\caption{
(a) Structure of Weyl photonic crystal. The locations of four air spheres with a radius of $0.07a$ in the double-gyroid unit cell are same with Ref.~\cite{yang2018ideal}. (b) Dispersion relation on the plane of $k_z=0$. The momentum $k_{x,y}$ is normalized by $2\pi/a$. (c) The real part of the radiative interaction $\Gamma_\mathrm{Re}$ (normalized by $\Gamma_\mathrm{Re}$(R=0)) as a function of distance for TLS transition frequencies (upper) $\omega =0.5545$, (middle) $0.5520$ and (lower) $0.5512 [2\pi c/a]$, which are marked with white contours i, ii, and iii in (b),respectively. The inter-TLS direction is $\hat{\mathbf{R}}=\left(-1, 1, 1\right)/\sqrt{3}$. The dipole orientations are $\hat{\boldsymbol{\mathrm{\muup}}}_{1,2}=\left(-1, 1, 1\right)/\sqrt{3}$ and $\boldsymbol{\mathrm{\muup}}_1$ is fixed at central point of the unit cell. Green dashed curves are the envelops of the solid curves. Inset (i-iii) are the isosurfaces in momentum space. (d) The linear relationship between decay length ${\ell}_D$ and inverse size of isosurfaces $1 / \ \overline{q}$.
}
\label{fig:fig3}
\end{figure*}

The graphic illustration also indicates that the interaction range is \textit{inversely} proportional to the size of the isosurface in momentum space. A large inter-TLS distance $R$ on a large isosurface leads to a fast oscillating integrand on the isosurface that results in a small value of the integral. One way to counteract this effect is to substantially reduce the isosurface size. Small isosurfaces can save the integral from cancellation even for a fast-oscillating function. Figure 2c shows the real part of the integrand $\rho_\mathbf{k} e^{i\mathbf{k}\cdot \mathbf{R}}$ with a long inter-TLS distance $R=10\lambda$ on an isosurface that has a radius that is $0.03$ times that of the free-space isosurface. While the oscillation is still fast, the small isosurface cannot accommodate many oscillations, yielding a sizable value of the integral. Figure 2e shows that this strong interaction is sustained over a long distance if the isosurface is small. Specifically, for an isosurface with a radius of $q$, the real part of interaction $\Gamma_\mathrm{Re}$ scales as ${\sin{\left( qR \right)}}/{ qR }$. As the isosurface radius approaches zero $q\to 0$, the range becomes infinite. Here, we use a polarization factor $\rho_\mathbf{k}$ based on plane waves, which, although a simplification, is sufficient for estimating the scaling.

The size of isosurface is fixed in free space. But there are many structured photonic environments that offer smaller isosurfaces. Here, we use Weyl photonic crystals as an example to demonstrate the inverse relationship between the interaction range and the isosurface. Weyl photonic crystals~\cite{lu2013weyl,yang2018ideal} exhibit a conic dispersion relation in three-dimensional space, similar to Dirac dispersion relations in two-dimensional space. The isosurface gradually reduces to a point around the apex of the conic dispersion, i.e. the Weyl point. Observation of this small isosurface suggests that we could expect long-range interactions around isolated Weyl points.
Specifically, we consider a double gyroid structure described by $g\left(\boldsymbol{\mathrm{r}}\right)\mathrm{=}{\mathrm{sin} (2\pi x/a)\ }{\mathrm{cos} (2\pi y/a)\ }+{\mathrm{sin} (2\pi y/a)\ }{\mathrm{cos} (2\pi z/a)\ }+{\mathrm{sin} (2\pi z/a)\ }{\mathrm{cos} (2\pi x/a)\ }$, where $a$ is the lattice constant. A material with a dielectric constant $\varepsilon_r=13$ fills the regions defined by $|g\left(\boldsymbol{\mathrm{r}}\right)|>1.1$. Four air spheres are placed in the dielectric material as defects to break parity symmetry yielding two pairs of Weyl points at identical frequencies~\cite{wang2016topological}. The unit structure is shown Fig. 3(a). The dispersion relation on the momentum plane of $k_z=0$ is shown in Fig. 3b with two pairs of Weyl points at the frequency ${\omega}_{\mathrm{wp}}=0.55096 [2\pi c/a]$. The isosurface becomes infinitesimally small at the Weyl point.

Using these isosurfaces, we numerically calculate the interaction between two TLSs placed inside the Weyl crystal. The photonic modes are simulated using the MPB software package~\cite{johnson2001block}. The details of the calculation are shown in SM. Figure 3c shows the interaction as a function of the inter-TLS distance for three different transition frequencies, which are also labeled by white lines in Fig. 3b.
The isosurfaces have four lobes because there are four Weyl points, as shown in Fig. 3c (i-iii). As the TLS transition frequency approaches the Weyl point, the isosurface size decreases, causing the interaction extends to extend to a significantly greater range. When the transition frequency is $0.00024 [2\pi c/a]$ away from the Weyl point (panel iii in Fig. 3c), the interaction shows a negligible decay even at $180$ wavelengths (Fig. 3c bottom).

The decaying and oscillating patterns in these curves are attributed to a few different origins. At the largest scale, the envelop scales as $\sin{ \left( \bar{q}R\right)}/ \bar{q}R$, where we use $\bar{q}$ to roughly characterize the size of the isosurface (we will discuss the impact of the shape of isosurface later). The medium-range oscillation is due to the interplay of four Weyl points at the same frequency. The fastest oscillation is due to the modulation of the nonuniform field within a unit cell of the photonic crystal. The long-range interaction observed here is robust in that it does not rely on the orientation of the dipole direction or the spatial placement of TLSs (See more discussion in SM).

We can quantitatively characterize the interaction range by numerically fitting the envelope. These envelops are shown by the dashed line in Fig. 3c. We further define a range $\ell_D$ as the distance when the envelop drops to half of its maximum value. We calculate this range for TLSs at different transition frequencies near the Weyl points, corresponding to different isosurface sizes. The results are shown in Fig. 3d. A clear linear relationship is demonstrated between $\ell_D$ and the inverse of the isosurface size $1/\bar{q}$. Because the isosurfaces are not spherical, we use $\bar{q}=\sqrt{S_{\omega}/4\pi}$ to define the isosurface size, where $S_{\omega}$ is the surface area of isosurfaces.

Thus far, we have shown that the size of the isosurface plays a critical role in the interaction range. Next, we will discuss the role of the shape of the isosurface. A spherical isosurface leads to an isotropic interaction range. On the other hand, a non-spherical isosurface generally creates an anisotropic interaction range: the interaction range depends on the direction of the inter-TLS distance vector $\hat{\mathbf{R}}$. There is a general reciprocal relationship between the interaction range and the size of the isosurface when projected along $\hat{\mathbf{R}}$.

Let us take the example of an ellipsoidal isosurface in an anisotropic media. The interaction range is longer when the two TLSs are placed along the direction of the short axis of the ellipsoid $\hat{\boldsymbol{s}}$, than when they are along the long axis $\hat{\boldsymbol{l}}$. We can easily see this effect by observing the oscillation pattern of $\rho_\mathbf{k} e^{i\mathbf{k}\cdot \mathbf{R}}$ on an ellipsoidal isosurface as shown in Fig. 4a. When $\hat{\mathbf{R}}$ is parallel to the long axis $\hat{\boldsymbol{l}}$, we have many oscillations and strong cancellation of the integration. On the other hand, when $\hat{\mathbf{R}}$ is parallel to the short axis $\hat{\boldsymbol{s}}$, we have fewer oscillations and weaker cancellation.

To demonstrate this effect in Weyl photonic crystals, we plot the isosurface at frequency $\omega ={\omega}_{\mathrm{wp}}+0.00404 [2\pi c/a]$, where the isosurface has a flat edge-softened rectangular geometry (Fig. 4b). We plot the real part of the integrand in Eq. (2) on the isosurface for three different ${\mathbf{R}}$. Here the magnitude of $\mathbf{R}$ is fixed, but its direction $\hat{\mathbf{R}}$ varies from the short axis ~$\hat{\boldsymbol{s}}$ to the long axis $\hat{\boldsymbol{l}}$. The cancellation effect is weaker when $\mathbf{R}$ is aligned with the short axis and stronger along the long axis. We also calculate the interaction as a function of the distance for the three directions shown in Fig. 4b. The range is conspicuously longer for TLSs placed along the short axis of the isosurface than that for the long axis as shown in Fig. 4c.
In the case shown in Fig. 4, the frequency is greatly detuned from the Weyl point, and thus, the interaction range is not as long as those shown in Fig. 3.

\begin{figure}
\begin{center}
\epsfig{figure=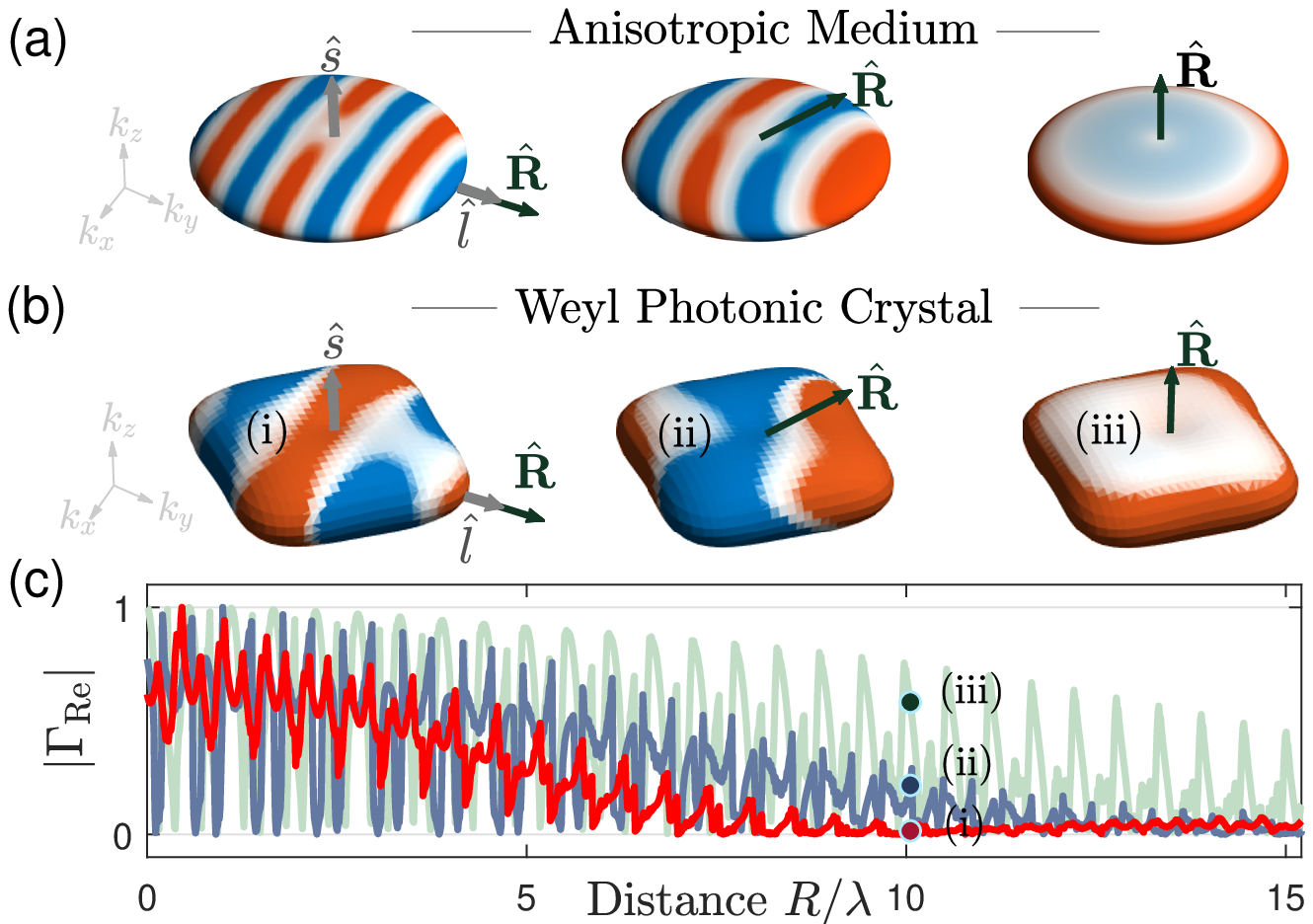,width=\linewidth}
\end{center}
\caption{(a) Real part of the integrand in Eq. (2) on an elliptical isosurface with (left) $\hat{\boldsymbol{\mathrm{R}}}\mathrm{=}\left(0,1,0\right)$, (middle) $\left(0,1,1\right)/ \sqrt{2}$, and (right) $\left(0,0,1\right).$ Unit vectors ~$\hat{\boldsymbol{s}}$ and ~$\hat{\boldsymbol{l}}$ represent short and long axis of the anisotropic isosurface. The dipole orientation is fixed as ~${\hat{\boldsymbol{\mathrm{\muup}}}}_{1,2}=\left(0,0,1\right)$. (b) Same as (a), but the isosurface is in the Weyl photonic crystal in Fig. 3a at frequency $\omega =0.555 [2\pi c/a]$ and the dipole orientation is fixed as ~${\hat{\boldsymbol{\mathrm{\muup }}}}_{1,2}=\left(0,1,0\right)$. (c) The absolute value of ${\mathrm{\Gamma }}_{\mathrm{Re}}$ as a function of distance $R$. Light green, blue and red curves, respectively, correspond to $\hat{\boldsymbol{\mathrm{R}}}$ in left, middle, and right cases of (b).}
\label{fig:fig4}
\end{figure}

The extended range of the dipole-dipole interaction extends beyond quantum systems. In the microwave regime, where Weyl photonic crystals have been experimentally realized on a printed circuit board~\cite{yang2018ideal}, the resonant dipole-dipole interaction range can also be extended. The range will also be limited by the propagation length of the waves inside such systems due to finite absorption by metallic materials.

We also emphasize that the relation between the interaction range and the isosurface is not unique to Weyl photonic crystals. It is generally applicable to periodically structured media.  For example, in two-dimensional space, the scaling of the interaction range follows $J_0(kR)$, where $J_0$ is the Bessel function of the first kind. For a two-dimensional photonic crystal,  a spherical isosurface with a radius of $q$ creates a different scaling law that follows $J_0(qR)$. More examples are discussed in Sec. II of SM.

We have discussed that the interaction range. Another important aspect is the strength of the interaction. We chose the linear dispersion near Weyl points because it makes it easy to separate the effect of the isosurface from other effects such as group velocity and density of states. However, the shrinking isosurface combined with a finite group velocity also decreases the interaction strength. At the Weyl point, the interaction strength is zero. The linear dispersion near a Weyl point results in a trade-off between the interaction range and strength. Such a trade-off can be alleviated in two-dimensional crystals and with a high order dispersion relation. We discussed the scaling of the interaction strength in Sec. II in SM.

Visual inspection of the isosurface provides a convenient tool to understand a broad class of long-range interaction phenomena. We now comment on the connection between our approach and the existing literature. The behavior of index-near-zero materials~\cite{fleury2013enhanced} was explained by a long effective wavelength. Alternatively, it can also be conveniently explained by our method: the index-near-zero material also has an ultra-small isosurface.
In addition to these examples, we can envision that Dirac points in two-dimensional photonic crystals also provide small `isosurfaces' (isofrequency contours) for long-range interaction.
Ref.~\cite{douglas2015quantum} shows that \textit{inside} the photonic bandgap, long tails of evanescent fields can induce long-range interaction. Here we can also see that \textit{outside} the photonic bandgap but near the band edge, the propagating far fields have small isosurfaces, offering a different mechanism for long-range interaction.
A hyperbolic material, where long-range interactions were allowed along specific directions, was treated using the Green's function method~\cite{cortes2017super}.
Using our graphic interpretation allows one to intuitively see that only special directions allow long-range interactions (see the visualization in SM).

To conclude, we show the deep connection between the interaction range and the isosurface in momentum space. Both the size and shape of the isosurface affect the interaction range. The method introduced here provides an intuitive understanding of underlying physics that is somewhat buried in traditional treatments, and we were able to use our method to help understand several photonic systems from the existing literature. It also provides a general recipe to search for new photonic systems that support long-range interactions.

This work was supported by the National Science Foundation (NSF) through the University of Wisconsin Materials Research Science and Engineering Center DMR-1720415. L.Y. and Z.Y. were also supported by the Defense Advanced Research Projects Agency (DARPA) (DETECT program). L. Y. also acknowledges the financial support from NSF EFRI Award-1641109.


%

\beginsupplement

\part*{Supplementary Material}

\section{Theory of long-range interaction}\label{sec:general_theory}

\subsection{General resonant interaction theory between two quantum two-level systems}

The Hamiltonian of quantum two-level systems (TLSs) in an arbitrary photonic environment is given by
\begin{equation}\label{eq:hamiltonian_tot}
   \mathcal{H} = \mathcal{H}_\mathrm{ph}+\mathcal{H}_\mathrm{tls}+\mathcal{H}_\mathrm{int}.
\end{equation}
They are explicitly written as~\cite{bay1997atom} ($\hbar=1$)
\begin{equation}
\begin{split}
   \mathcal{H}_\mathrm{ph} = &\sum_{\mathbf{k},\eta}\omega_\mathbf{k} \hat{a}_{\mathbf{k},\eta}^{\dag} \hat{a}_{\mathbf{k},\eta} \\
   \mathcal{H}_\mathrm{tls} = & \sum_{i=1,2}\omega_0 \hat{\sigma}_{i}^{\dag} \hat{\sigma}_{i} \\
   \mathcal{H}_\mathrm{int} = & \sum_{i=1,2}\sum_{\mathbf{k},\eta}
   \Big[ i g_{\mathbf{k},\eta}(\mathbf{r}_i)
   \left(\hat{\sigma}_{i}^{\dag} + \hat{\sigma}_{i}\right) \hat{a}_{\mathbf{k},\eta}e^{i\mathbf{k}\cdot \mathbf{r}_i} + \mathrm{H.c.} \Big],
\end{split}
\end{equation}
where $\omega_0$ and $\hat{\sigma}_i^\dag (\hat{\sigma}_i)$ are the transition frequency and raising (lowering) operator of $i$th two-level system (TLS). $\omega_\mathbf{k}$ and $\hat{a}_{\mathbf{k},\eta}^{\dag} (\hat{a}_{\mathbf{k},\eta})$ are the frequency and creation (annihilation) operator of photon. $g_{\mathbf{k},\eta}(\mathbf{r}_i)=\sqrt{\omega_\mathbf{k}/2\varepsilon_0 V} \boldsymbol{\mathrm{\muup}}_i \cdot \boldsymbol{\mathrm{\epsilonup}}_{\mathbf{k},\eta} $ is the coupling between $i$th TLS and photonic mode $\mathbf{k}$. $\boldsymbol{\mathrm{\muup}}_i$ is the transition dipole moment of $i$th quantum TLS and $\boldsymbol{\mathrm{\epsilonup}}_{\mathbf{k},\eta}$ is the polarization of photonic mode $\mathbf{k}$ with polarization index $\eta$.

The transition probability from initial to final states is given by the Fermi's Golden rule
$ 2\pi / \hbar |\mathcal{M}_\mathrm{F I}|^2
\delta\left( \mathcal{E}_\mathrm{F}-\mathcal{E}_\mathrm{I} \right)$,
where the transition matrix element $\mathcal{M}_\mathrm{F I}$ can describe the resonant dipole-dipole interaction between two TLSs.
For the weak light-matter interaction, it can be written as the second-order form:
\begin{equation}
\mathcal{M}_\mathrm{F I} =  \langle \mathrm{F}| \mathcal{H}_{\mathrm{int}} | \mathrm{I} \rangle + \sum_\alpha \frac{ \langle \mathrm{F}| \mathcal{H}_{\mathrm{int}} | \mathrm{R}_\alpha \rangle  \langle \mathrm{R}_\alpha | \mathcal{H}_{\mathrm{int}} | \mathrm{I} \rangle   }{\mathcal{E}_\mathrm{I} -\mathcal{E}_{\mathrm{R}_\alpha} } + \cdots.
\end{equation}
Here, $|\mathrm{I}\rangle = | e_1,g_2;0\rangle$ and $|\mathrm{F}\rangle = | g_1,e_2;0\rangle$ denote initial and final states, where`$e$' and `$g$' in the Dirac bracket notions represent excited and ground states, respectively, and the number `$0$' or `$1$' is the photon number in the photonic environment. The intermediate state $| \mathrm{R}_\alpha \rangle$ has two options: $|g_1,g_2;1_{\mathbf{k},\eta}\rangle$ with energy $\mathcal{E}_{\mathrm{R}_1 }=\mathcal{E}_g^{(1)}+\mathcal{E}_g^{(2)}+\hbar \omega_\mathbf{k}$ and $| e_1,e_2;1_{\mathbf{k},\eta}\rangle$ with energy $\mathcal{E}_{\mathrm{R}_2 }=\mathcal{E}_e^{(1)}+\mathcal{E}_e^{(2)}+\hbar \omega_\mathbf{k}$. The energy of the initial state is $\mathcal{E}_{\mathrm{I} }=\mathcal{E}_g^{(1)}+\mathcal{E}_g^{(2)}+\hbar \omega_\mathbf{k}$. Since two identical TLSs are considered, we have $\mathcal{E}_e^{(1,2)}-\mathcal{E}_g^{(1,2)}=\hbar \omega_0$. Then, Eq. (3) can be explicitly given by
\begin{equation}
\begin{split}
\mathcal{M}_\mathrm{FI}=
\sum_{{\mathbf{k}},\eta}
\Big(
& g_{\mathbf{k},\eta}\left(\mathbf{r}_1\right)^\ast g_{\mathbf{k},\eta} \left(\mathbf{r}_2\right)
 \frac{e^{i\mathbf{k}\cdot \mathbf{R} } }{ \omega_\mathbf{k} -\omega_0 }  \\
 +
& g_{\mathbf{k},\eta}\left(\mathbf{r}_1\right) g_{\mathbf{k},\eta}\left(\mathbf{r}_2\right)^\ast 
\frac{e^{-i\mathbf{k}\cdot \mathbf{R} } }{ \omega_\mathbf{k} + \omega_0 } \Big) ,
\end{split}
\end{equation}
where $\mathbf{R}=\mathbf{r}_2 - \mathbf{r}_1$ and $\omega_\mathbf{k}=\omega\left(\mathbf{k}\right)$. The summation over
$\mathbf{k}$ can be written as an integral as $\sum_{\boldsymbol{\mathrm{k}}}= [{V}/{{\left(2\pi \right)}^3} ]  \iiint_{\mathcal{V}_\mathbf{k}} d \mathbf{k}^3$.
The transition matrix element can also be expressed as~\cite{kurizki1990two}$\mathcal{M}_\mathrm{FI}=i\Gamma^\ast$, where $\Gamma$ is the radiative interaction.
Utilizing the relation $\int_0^\infty [f(x)/\left(x-x_0\right)] dx=\int_0^\infty \left[ \mathbb{P}\left( f(x)/\left(x-x_0\right) \right) + i\pi\delta(x-x_0)f(x) \right] dx$, we can write the radiative interaction as
\begin{equation}
\begin{split}
\Gamma = &
 \ \Gamma_\mathrm{Re}+i\Gamma_\mathrm{Im} \\
 \\
= &  \sum_\eta \iiint_{ \mathcal{V}_\mathbf{k} } d^3\mathbf{k}
\frac{ v_g(\mathbf{k})}{\pi }
\Bigg[
 \pi \delta \left( \omega_\mathbf{k} -\omega_0 \right) \rho_{\mathbf{k},\eta} e^{i\boldsymbol{\mathrm{k}}\cdot {{\mathbf{R}}}}
 \\
& \ \ \ \ \ \ \ \ \ +
i\mathbb{P}
\left(
\rho_{\mathbf{k},\eta}
\frac{ e^{i\boldsymbol{\mathrm{k}}\cdot {{\mathbf{R}}}}}{\omega_\mathbf{k}-\omega_0 }
+
\rho_{\mathbf{k},\eta}^\ast
\frac{e^{-i\mathbf{k} \cdot \mathbf{R}}}{\omega_\mathbf{k}+\omega_0}
\right)
 \Bigg],
\end{split}
\end{equation}
where $\mathbb{P}$ denotes the Cauchy principal value
and the polarization factor is
\begin{equation}
\rho_{\mathbf{k},\eta}=\frac{\omega}{16\pi^2\varepsilon_0 v_g(\mathbf{k})} \left( \boldsymbol{\mathrm{\muup}}_1\cdot \boldsymbol{\mathrm{\epsilonup}}_{\mathbf{k},\eta} \right)^\ast
\left(
\boldsymbol{\mathrm{\muup}}_2 \cdot \boldsymbol{\mathrm{\epsilonup}}_{\mathbf{k},\eta} \right).
\end{equation}
The real part of $\Gamma$ is the cooperative decay rate and its explicit expression is
\begin{equation}\label{eq:gamma_re}
\Gamma_\mathrm{Re} \left(\omega_0\right) =
 \sum_\eta \iint_{\mathcal{S}_{\omega_0(\mathbf{k})}}
\rho_{\mathbf{k},\eta}\left( \omega_0 \right)
{e^{i\mathbf{k} \cdot \mathbf{R} } }
d\mathcal{S}_\mathbf{k},
\end{equation}
where $\mathcal{S}_{\omega_0(\mathbf{k})}$ is the isosurface of $\omega=\omega_0$ in momentum space and $v_g\left(\mathbf{k}\right)
=|\nabla_\mathbf{k} \omega_\mathbf{k} |$ is the group velocity of mode $\mathbf{k}$.
The cooperative energy shift is
\begin{equation}\label{eq:energ_shift}
\begin{split}
\Gamma_\mathrm{Im} \left(  \omega_0 \right) =&   \frac{1}{\pi  }
\mathbb{P}\int_0^{\infty}  d\omega
\sum_\eta
\iint_{\mathcal{S}_{\omega(\mathbf{k})}} dS_\mathbf{k}
  \\
& \ \      \times
   \left( \frac{\rho_{\mathbf{k},\eta} }{\omega-\omega_0}  e^{i\boldsymbol{\mathrm{k}}\cdot \mathbf{R} }
 +  \frac{\rho_{\mathbf{k},\eta}^\ast}{\omega+\omega_0}   e^{-i\boldsymbol{\mathrm{k}}\cdot \mathbf{R} }  \right)  \\
 \\
= &   \frac{1}{\pi }
\mathbb{P}\int_0^{\infty}  d\omega
\left(
\frac{\Gamma_\mathrm{Re}(\omega)}{\omega-\omega_0}
 +  \frac{\Gamma_\mathrm{Re}^\ast(\omega)}{\omega+\omega_0}
 \right).
\end{split}
\end{equation}

\subsection{Interaction in free space vacuum}

 In the free space, the dispersion relation is given by
\begin{equation}
\omega_\mathbf{k}\ = c|\mathbf{k}| = ck ,
\end{equation}
where $c$ is the speed of light. Because the isosurface is isotropic, we have
$\iint_{\mathcal{S}_{\omega(\mathbf{k})}} d\mathcal{S}_\mathbf{k} = k^2 \iint d\Omega_\mathbf{k} = k^2\int^{\mathrm{2}\pi}_0 d\theta \int^{\pi}_0 d\varphi \mathrm{sin}\varphi$ and group velocity $v_g\left(\mathbf{k}\right)=c$.
Assuming $\boldsymbol{\mathrm{k}}\mathrm{\cdot }{\boldsymbol{\mathrm{R}}}\mathrm{=}{\xi }_{\boldsymbol{\mathrm{k}}}kR =  \mathrm{cos}\varphi kR$, the cooperative decay rate in free space is written as ($c=1$)
\begin{equation}
\begin{split}
\Gamma_\mathrm{Re} = & \frac{\mu_1 \mu_2 k^3}{16\pi^2\varepsilon_0 } \sum_\eta \int_0^{2\pi}d\theta \int_0^{\pi} d\varphi \sin{\varphi}\\
\times &
\left( \boldsymbol{\mathrm{\muup}}_1\cdot \boldsymbol{\mathrm{\epsilonup}}_{\mathbf{k},\eta} \right)^\ast \left( \boldsymbol{\mathrm{\muup}}_2 \cdot \boldsymbol{\mathrm{\epsilonup}}_{\mathbf{k},\eta} \right)
e^{ik R \xi_\mathbf{k} }.
\end{split}
\end{equation}
The polarization sum rule is given by
\begin{equation}\label{eq:sum_rule}
\sum_{\eta} {\epsilon^{(\eta)}_{\mathbf{k},i} } {{\epsilon}^{(\eta)}_{\mathbf{k},j}} = {\delta}_{12}\mathrm{-}{\hat{k}}_1 {\hat{k}}_2
\end{equation}
with $\epsilon^{(\eta)}_{\mathbf{k},i} = \hat{\boldsymbol{\mathrm{\muup}}}_i\cdot \boldsymbol{\mathrm{\epsilonup}}_{\mathbf{k},\eta}$, $\delta_{12} = \hat{\boldsymbol{\mathrm{\muup}}}_1 \cdot \hat{\boldsymbol{\mathrm{\muup}}}_2 $, and ${\hat{k}}_i= \hat{\mathbf{k}} \cdot \hat{\boldsymbol{\mathrm{\muup}}}_i $.
Then, we have
\begin{equation}
\begin{split}
\Gamma_\mathrm{Re} = & \frac{\mu_1 \mu_2 k^3}{16\pi^2\varepsilon_0} \int_0^{2\pi}d\theta \int_0^{\pi} d\varphi
\left( \delta_{12} - {\hat{k}}_1 {\hat{k}}_2 \right)
\sin{\varphi} e^{ik R  \xi_\mathbf{k} }   \\
= & \frac{\mu_1 \mu_2 k }{16\pi^2\varepsilon_0}
\left( -\nabla^2\delta_{12} + \nabla_1\nabla_2 \right)
\int_0^{2\pi}d\theta \int_0^{\pi} d\varphi
\sin{\varphi}e^{ik R \xi_\mathbf{k} } \\
= & \frac{\mu_1 \mu_2 k }{4\pi\varepsilon_0}
\left( -\nabla^2\delta_{12} + \nabla_1\nabla_2 \right)
\frac{\sin{k R} }{ R } \\
= & \frac{\mu_1 \mu_2 k^3 }{4\pi\varepsilon_0}
\Bigg(
\left( \delta_{12}- \hat{{R}}_{i}  \hat{{R}}_{j} \right)
\frac{\sin{k R} }{k R} \\
 & \ \ \ \ \ \ \ \ \ \ +
\left( \delta_{12}- 3 \hat{{R}}_{i} \hat{{R}}_{j} \right)
\left(
\frac{\cos{k R} }{(k R)^2} +\frac{\sin{k R} }{(k R)^3} \right)
\Bigg),
\end{split}
\end{equation}
where $\hat{{R}}_{1,2} = \hat{\boldsymbol{\mathrm{\muup}}}_{1,2} \cdot \hat{\mathbf{R}} $.
Also, with the relation in Eq.~(\ref{eq:sum_rule}), the cooperative energy shift is given by
\begin{equation}
\begin{split}
\Gamma_\mathrm{Im}
= &
\frac{\mu_1 \mu_2 }{16\pi^3\varepsilon_0 }
\left( -\nabla^2\delta_{12} + \nabla_1\nabla_2 \right)
\int_0^{2\pi}d\theta \int_0^{\pi} d\varphi \sin{\varphi}  \\
& \times
\mathbb{P}\int_{0}^\infty dk k
\left(
\frac{e^{ikR\xi_\mathbf{k}}}{k-k_0} + \frac{e^{-ikR\xi_\mathbf{k}}}{k+k_0}
\right) .
\end{split}
\end{equation}
After calculating the Cauchy principal integral and integral over isosurface~\cite{andrews2004virtual}, we have
\begin{equation}
\begin{split}
\Gamma_\mathrm{Im}
= &
\frac{\mu_1 \mu_2 k_0}{4\pi \varepsilon_0}
\left( -\nabla^2\delta_{12} + \nabla_1\nabla_2 \right)
\frac{\cos{k_0R}}{R}\\
= &
\frac{\mu_1 \mu_2 k_0^3 }{4\pi\varepsilon_0 }
\Bigg(
\left( \delta_{12}- \hat{{R}}_{i}  \hat{{R}}_{j} \right)
-\frac{\cos{k_0 R} }{k_0 R} \\
 & \ \ \ \ \ +
\left( \delta_{12}- 3 \hat{{R}}_{i} \hat{{R}}_{j} \right)
\left(
\frac{\sin{k_0 R} }{(k_0 R)^2} +\frac{\cos{k_0 R} }{(k_0 R)^3} \right)
\Bigg).
\end{split}
\end{equation}

\subsection{Interaction near Weyl points}
At first, we only consider a single Weyl point, as shown in Fig. 2c of the main text.
The Hamiltonian for the continuum around the Weyl point is given by~\cite{lu2013weyl}
\begin{equation}
\mathcal{H}_\mathrm{wp}\left( \mathbf{k} \right) = \sum_{i=x,y,z} v_i q_i \sigma_i,
\end{equation}
where $\sigma_{x,y,z}$ are Pauli matrices and $\mathbf{q}=\left( q_x,\ q_y,\ q_z \right)=\mathbf{k}-\mathbf{k}_c$ is the distance to the Weyl point in momentum space. The Weyl point is at $\mathbf{k}_c$ when $\mathbf{q}=0$. $v_{x,y,z}$ are the $x,y,z$ components of the group velocity. For simplicity, we assume the isosurface is isotropic, i.e. $v_x=v_y=v_z=v$. The dispersion relation near the Weyl point is given by
\begin{equation}\label{eq:weyl_dispersion}
    \omega_\mathbf{q} = \omega_\mathrm{wp} \pm  v |\mathbf{q}|  .
\end{equation}
Then, the cooperative decay rate is given by
\begin{equation}
\begin{split}
\Gamma_\mathrm{Re}(\omega) = &
\frac{\mu_1 \mu_2 \omega}{16\pi^2\varepsilon_0 v}
e^{i \mathbf{k}_\mathrm{c} \cdot \mathbf{R}} \\
&\ \ \ \ \  \times
\iint_{\mathcal{S}_{\omega  \left( \mathbf{q}\right) }} d\mathcal{S}_\mathbf{q}
\left( \hat{\boldsymbol{\mathrm{\muup}}}_1\cdot \boldsymbol{\mathrm{\epsilonup}}_{\mathbf{k},\eta} \right)^\ast \left( \boldsymbol{\hat{\mathrm{\muup}}}_2 \cdot \boldsymbol{\mathrm{\epsilonup}}_{\mathbf{k},\eta} \right)
e^{i \mathbf{q}\cdot \mathbf{R}}.
\end{split}
\end{equation}
Here, we use a polarization based on plane waves. Although this approximation is simplified, it is sufficient for calculating the scaling.
At frequencies near the Weyl point, $q\ll k_\mathrm{c}$ and the polarization factor term is a constant
\begin{equation}
\rho_{\mathbf{k},\eta} =  \frac{\mu_1 \mu_2 \omega}{16\pi^2\varepsilon_0 v}
\left( \hat{\boldsymbol{\mathrm{\muup}}}_1\cdot \boldsymbol{\mathrm{\epsilonup}}_{\mathbf{k},\eta} \right)^\ast \left( \boldsymbol{\hat{\mathrm{\muup}}}_2 \cdot \boldsymbol{\mathrm{\epsilonup}}_{\mathbf{k},\eta} \right)
\simeq \rho.
\end{equation}
Consequently, the cooperative decay rate is
\begin{equation}
\begin{split}
\Gamma_\mathrm{Re}
\simeq &
 \rho
e^{i \mathbf{k}_\mathrm{c} \cdot \mathbf{R}}
\iint_{\mathcal{S}_{\omega_0 \left( \mathbf{q}\right) }} d\mathcal{S}_\mathbf{q}
e^{i\mathbf{q} \cdot \mathbf{R} } \\
= &
4\pi  q^2 \rho
e^{i \mathbf{k}_\mathrm{c} \cdot \mathbf{R}}
\frac{\sin{qR}}{qR} .
\end{split}
\end{equation}

The imaginary part of the radiative interaction (the cooperative energy shift) is an integral over frequencies from zero to infinity, as shown in Eq. ~(\ref{eq:energ_shift}). For Weyl photonic crystals, the dispersion relation is quite different from Eq.~(\ref{eq:weyl_dispersion}) at frequencies far from the Weyl point~\cite{lu2013weyl}. Thus, we do not show an analytic estimation of the cooperative energy shift here, but numerical details will be discussed in Sec.~\ref{sec:Weyl_PC}.

{
\section{The scaling of the interaction strength in 3D and 2D photonic environments}

The interaction range increases as the isosurface decreases. However, at the same time, the density of states also decreases, particularly when the group velocity does not scale to zero. The consequence is that the interaction strength reduces while the range extends unless the group velocity scales in a way to cancel the effect. Here below, we discuss how the strength scales in different photonic environments.

\subsection{Three-dimensional photonic media}

In the 3D case, the derivation starts from the definition of the real part of ${\Gamma}$ as shown in Eq.~(\ref{eq:gamma_re}).  Here, we will show the real part of radiative interaction in different 3D photonic media.

\subsubsection{3D free space}

For the free space case, the dispersion relation is ${\omega }_{\boldsymbol{\mathrm{k}}}=c\left|\boldsymbol{\mathrm{k}}\right|=ck$, where $c$ is the speed of light. The radiative interaction strength is given by (see detailed in Sec.~\ref{sec:general_theory} B)
\begin{equation}
\begin{split}
\Gamma_\mathrm{Re}=
& B^{\left(3D\right)}\frac{k^2}{c}
\Bigg(
\left({\delta }_{12}-{\hat{R}}_1{\hat{R}}_2\right)\ \frac{{\mathrm{\ sin} kR\ }}{kR}+ \\
& \left(\delta_{12}-3{\hat{R}}_1{\hat{R}}_2\right)\ \left(\frac{{\mathrm{\ cos} kR\ }}{{\left(kR\right)}^2}+\frac{{\mathrm{\ sin} kR\ }}{{\left(kR\right)}^3}\right)\ \Bigg),
\end{split}
\end{equation}
where $B^{\left(3D\right)}={{\mu }_1{\mu }_2\omega }/{4\pi {\varepsilon }_0 }$.   In the long-distance regime, we have
\begin{equation}
\Gamma_{\mathrm{Re}}\left(R>\lambda \right)\cong B^{\left(3D\right)}\left({\delta }_{12}-{\hat{R}}_1{\hat{R}}_2\right)\frac{k^2}{c}\frac{{\mathrm{\ sin} kR\ }}{kR}.
\end{equation}

\subsubsection{3D Weyl photonic environment}

The linear dispersion relation near a Weyl point is given by Eq.~(\ref{eq:weyl_dispersion}).
Assuming $\left( \hat{\boldsymbol{\mathrm{\muup}}}_1\cdot \boldsymbol{\mathrm{\epsilonup}}_{\mathbf{k},\eta} \right)^\ast \left( \boldsymbol{\hat{\mathrm{\muup}}}_2 \cdot \boldsymbol{\mathrm{\epsilonup}}_{\mathbf{k},\eta} \right)\approx 1$, we have the radiative interaction (also see details in Sec.~\ref{sec:general_theory}C)
\begin{equation}
\begin{split}
{\mathrm{\Gamma }}_{\mathrm{Re}}
&\cong \frac{{\mu}_1{\mu}_2\omega}{16{\pi }^2{\varepsilon }_0  }\frac{1}{v}e^{i{\boldsymbol{\mathrm{k}}}_c\cdot \boldsymbol{\mathrm{R}}} \iint_{S_{\omega \left(\boldsymbol{\mathrm{q}}\right)}}{dS_{\boldsymbol{\mathrm{q}}}e^{i\boldsymbol{\mathrm{q}}\cdot \boldsymbol{\mathrm{R}}}}     \\
&=B^{\left(3D\right)}\frac{q^2\ }{v}\frac{{\mathrm{sin} qR\ }}{qR}e^{i{\boldsymbol{\mathrm{k}}}_c\cdot \boldsymbol{\mathrm{R}}}.
\end{split}
\end{equation}
As the radius of the small isosurface $q$ shrinks to zero, the interaction strength accordingly diminishes to zero. However, if the group velocity $v$ and the radius of isosurface $q$ are finite and small, the interaction can still be much larger and longer than that in vacuum (see comparison in Fig.~S1(d)).

\begin{figure}
\begin{center}\label{fig:large_strength}
\epsfig{figure=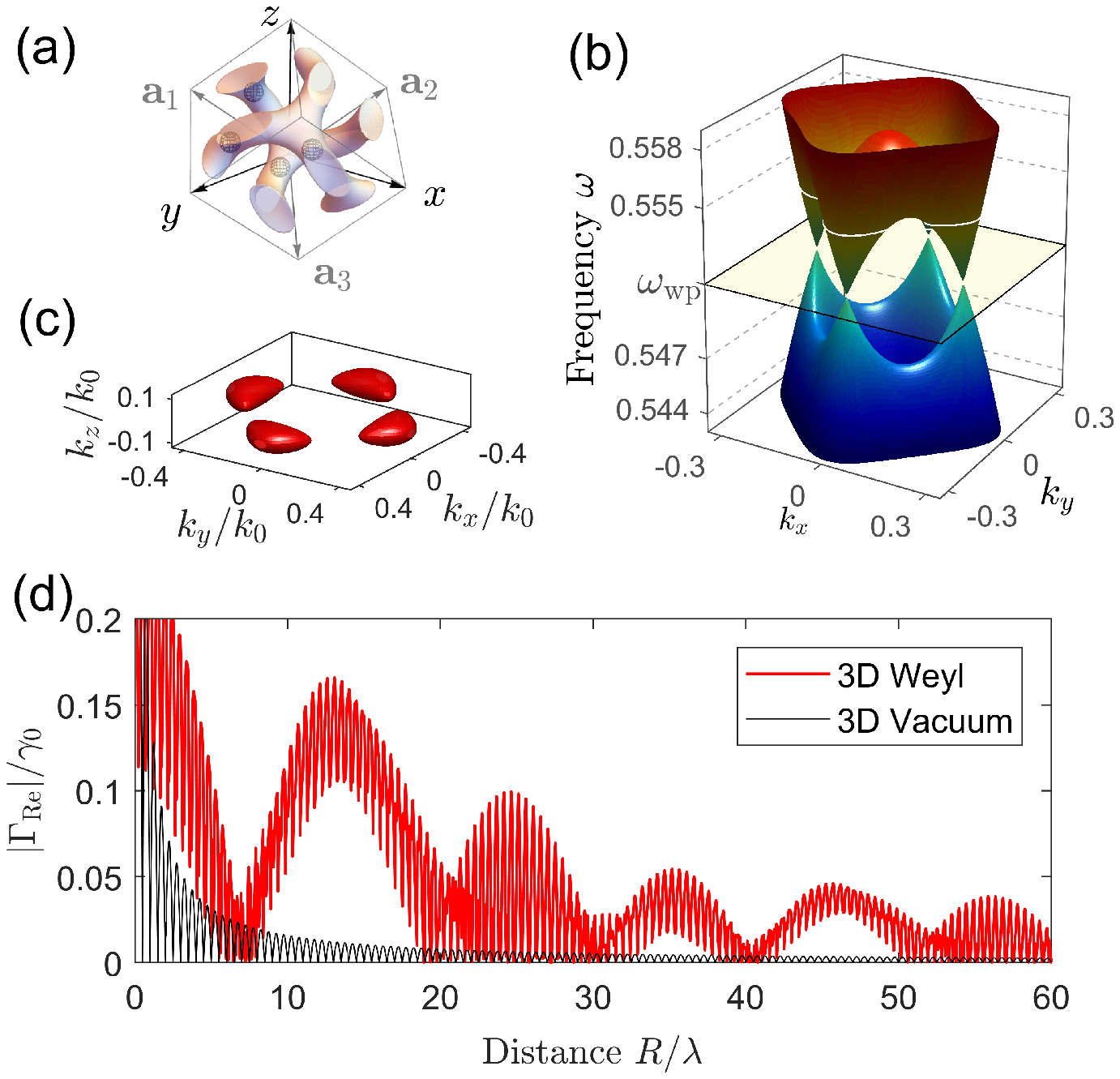,width=\linewidth}
\end{center}
\caption{
(a) Double-gyroid dielectric structure in a body-centered cubic unit cell with a set of basis vectors $\mathbf{a}_1=(-1/2,1/2,1/2)a$, $\mathbf{a}_2= (1/2,-1/2,1/2)a$, and $\mathbf{a}_3= (1/2,1/2,-1/2)a$. Four air spheres with a radius $r=0.07a$ are located at $(1/4, -1/8, 1/2)a$, $(1/4, 1/8, 0)a$, $(5/8, 0, 1/4)a$ and $(3/8, 1/2, 1/4)a$, respectively. The dielectric constant of solid gyroid structure is 13. (b) Dispersion relation on $k_z=0$ plane. The white curves correspond to isosurface in (c). (c) Isosurface at $\omega =0.5545 \ [2\pi c/a]$ normalized by the momentum in free space $k_0$. (d) The radiative interaction $\Gamma_{\mathrm{Re}}$ in vacuum and in the Weyl photonic crystal. The distance direction is $\hat{\boldsymbol{\mathrm{R}}}=[-1,1,1]$ and the dipole orientation is ${\hat{\boldsymbol{\mathrm{\muup}}}}_{1,2}=[0,0,1]$. $\gamma_0$ is the spontaneous decay rate in free space and $\lambda_0$ is the wavelength in free space.
}
\end{figure}

\subsubsection{3D quadratic photonic environment}

For a quadratic dispersion relation such as
\begin{equation}
{\omega}_{\mathbf{k}}=\beta {\left|\boldsymbol{\mathrm{k}}\boldsymbol{\mathrm{-}}{\boldsymbol{\mathrm{k}}}_{\boldsymbol{\mathrm{c}}}\right|}^2+{\omega }_c=\beta q^2+{\omega }_c,
\end{equation}
the group velocity is $v_g=2\beta q$. Supposing ${\boldsymbol{\mathrm{k}}}_{\boldsymbol{\mathrm{c}}}=0$, we have
\begin{equation}
\begin{split}
\Gamma_{\mathrm{Re}}
=&\frac{\omega }{16{\pi }^2{\varepsilon }_0 }\frac{q}{2\beta }\mathop{\int\!\!\!\!\int}\nolimits_{S_{\omega \left(\boldsymbol{\mathrm{q}}\right)}}{dS_{\boldsymbol{\mathrm{q}}}{\left({\boldsymbol{\mathrm{\muup }}}_1\cdot {\boldsymbol{\mathrm{\epsilonup }}}_{\boldsymbol{\mathrm{k}}}\right){\left({\boldsymbol{\mathrm{\muup }}}_2\cdot {\boldsymbol{\mathrm{\epsilonup }}}_{\boldsymbol{\mathrm{k}}}\right)}^{*\ }e}^{i\boldsymbol{\mathrm{q}}\cdot \boldsymbol{\mathrm{R}}}} \\
=&B^{\left(3D\right)}\frac{q}{2\beta }
\Bigg(
\left({\delta }_{12}-{\hat{R}}_1{\hat{R}}_2\right)\ \frac{{\mathrm{\ sin} qR}}{qR}+ \\
& \left(\delta_{12}-3{\hat{R}}_1{\hat{R}}_2\right)\ \left(\frac{{\mathrm{cos} qR\ }}{{\left(qR\right)}^2}+\frac{{\mathrm{sin} qR\ }}{{\left(qR\right)}^3}\right)\ \Bigg)
.
\end{split}
\end{equation}
In the long-distance regime, the radiative interaction is written as
\begin{equation}
\Gamma_{\mathrm{Re}}\left(R>\lambda \right)\cong B^{\left(3D\right)}\left({\delta }_{12}-{\hat{R}}_1{\hat{R}}_2\right)\frac{q}{2\beta }\frac{{\mathrm{sin} qR}}{qR}.
\end{equation}
The interaction strength reduces to zero when \textit{q} = 0, which is similar to Weyl point. However, when the group velocity $v$ and the radius of isosurface $q$ are finite and small, the radiative interaction can be significantly larger than that in vacuum.

\begin{figure}
\begin{center}
\epsfig{figure=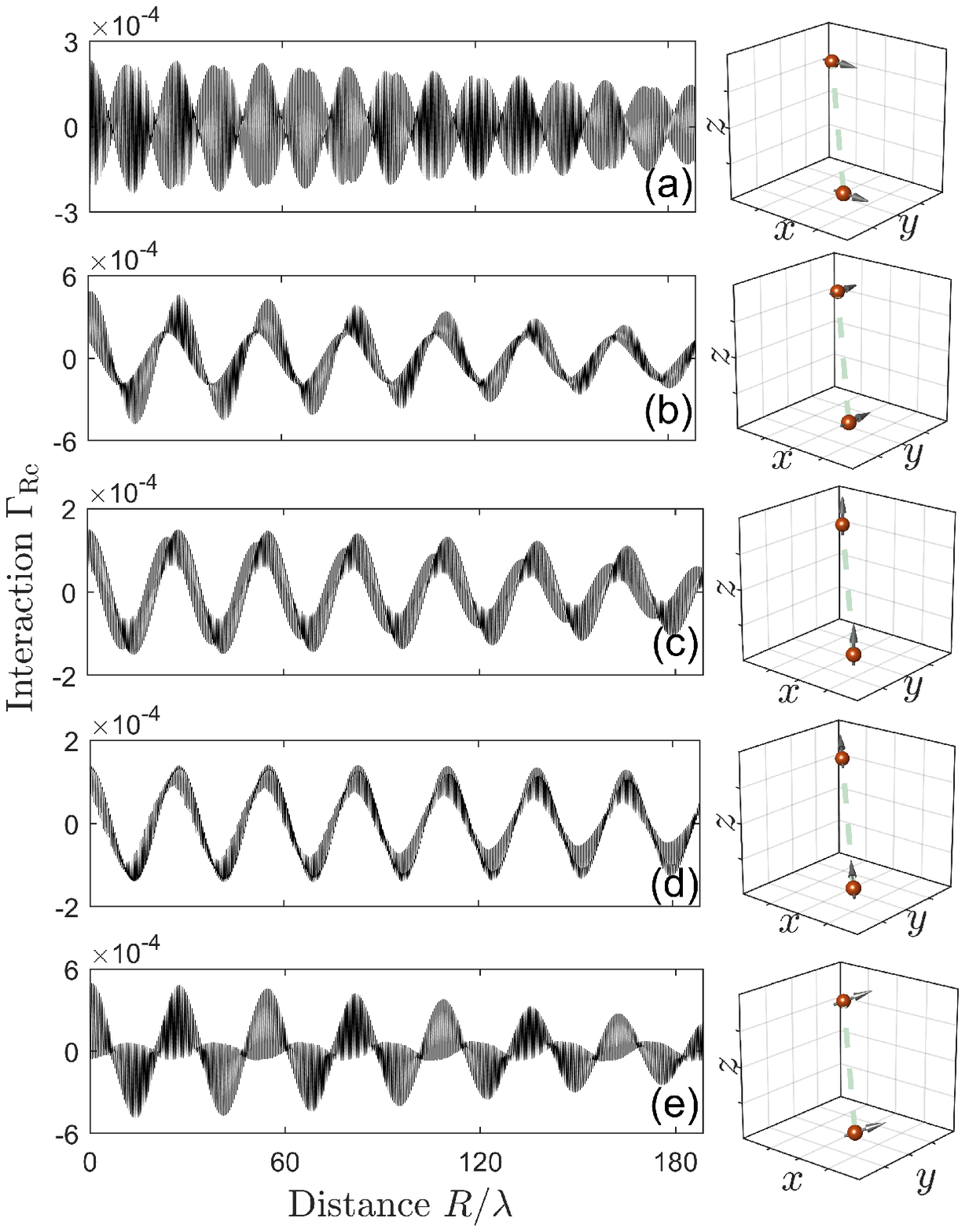,width=\linewidth}
\end{center}
\caption{
The real part of radiative interaction $\Gamma_\mathrm{Re}$ as a function of distance $R$ with dipole orientations (a) $\hat{\boldsymbol{\mathrm{\muup}}}_{1,2} =\left(1,\ 0,\ 0 \right)$, (b) $\left(0,\ 1,\ 0 \right)$,
(c) $\left(0,\ 0,\ 1 \right)$, (d) $\left(-1,\ 1,\ 1 \right)/\sqrt{3}$, and (e) $\left(2,\ 1,\ 1 \right)/\sqrt{6}$ at $\omega_0=0.5512 [2\pi c/a]$. The wavelength is given by $\lambda=2\pi c/\omega$. $\Gamma_\mathrm{Re}$ is normalized by ${\mu_1\mu_2 \omega_0 }/{16\pi^2\varepsilon_0}$.
}
\label{fig:dipole}
\end{figure}

\begin{figure}
\begin{center}
\epsfig{figure=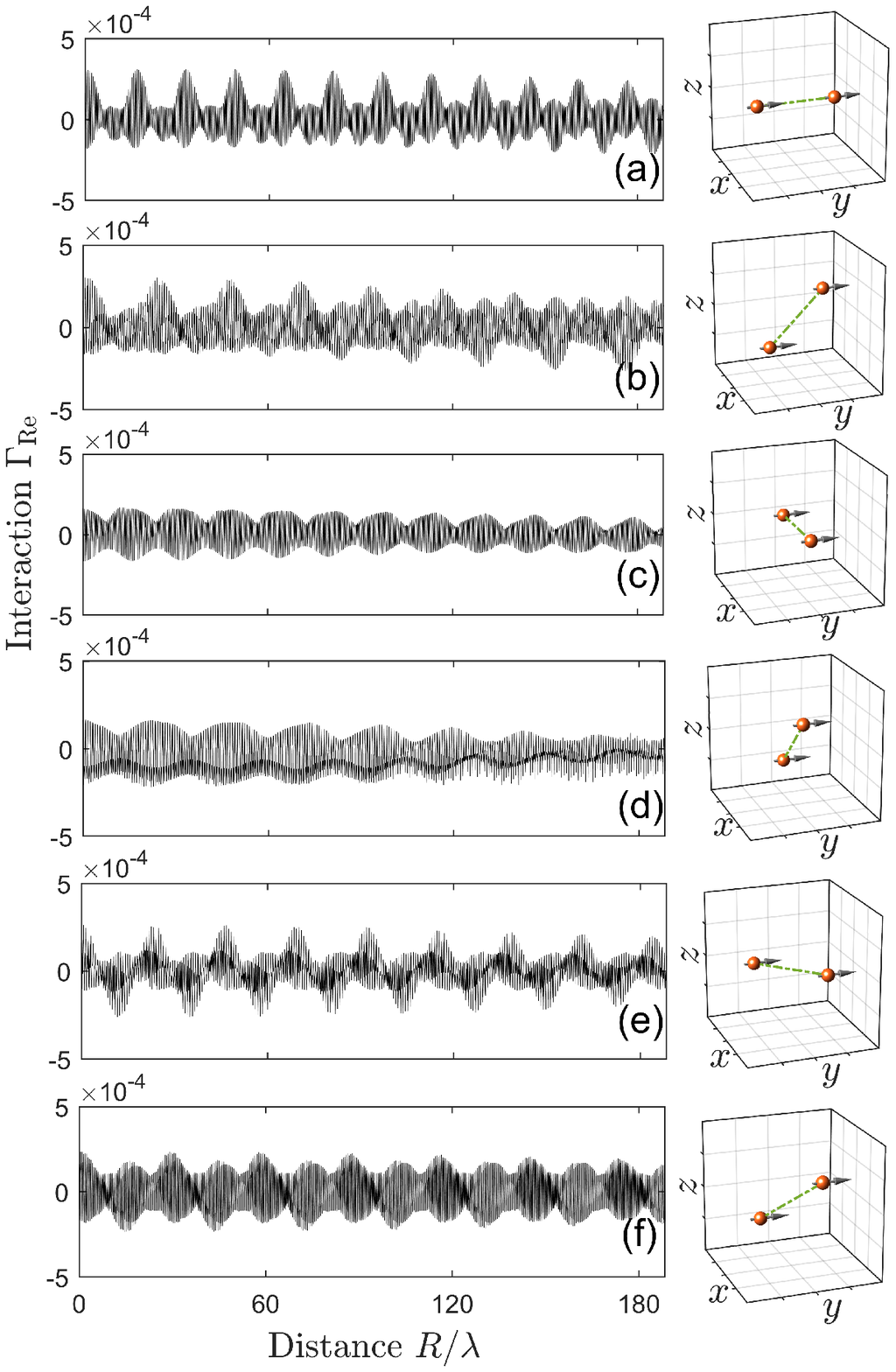,width=\linewidth}
\end{center}
\caption{
The real part of radiative interaction $\Gamma_\mathrm{Re}$ as a function of distance $R$ with distance vectors (a) $\hat{\mathbf{R}} =\left(0,\ 1,\ 0 \right)$, (b) $\left(0,\ 1,\ 1 \right)/\sqrt{2}$,
(c) $\left(1,\ 0,\ 0 \right)$, (d) $\left(1,\ 0,\ 1 \right)/\sqrt{2}$, (e) $\left(1,\ 1,\ 0 \right)/\sqrt{2}$, and (f) $\left(1,\ 1,\ 1 \right)/\sqrt{3}$ at $\omega_0=0.5512 [2\pi c/a]$. The wavelength is given by $\lambda=2\pi c/\omega$. $\Gamma_\mathrm{Re}$ is normalized by ${\mu_1\mu_2 \omega_0 }/{16\pi^2\varepsilon_0}$.
}
\label{fig:distance}
\end{figure}

\subsection{Two-dimensional photonic media }

In 2D case, the real part of the radiative interaction is written as
\begin{equation}
\Gamma_{\mathrm{Re}}\left(\omega \right)=\frac{\omega }{8\pi {\varepsilon }_0\hslash }\int_{\ell_{\omega \left(\boldsymbol{\mathrm{k}}\right)}}{d{\ell}_{\boldsymbol{\mathrm{k}}}}\left({\boldsymbol{\mathrm{\muup }}}_1\cdot {\boldsymbol{\mathrm{\epsilonup }}}_{\boldsymbol{\mathrm{k}}}\right){\left({\boldsymbol{\mathrm{\muup }}}_2\cdot {\boldsymbol{\mathrm{\epsilonup }}}_{\boldsymbol{\mathrm{k}}}\right)}^{*\ }\frac{e^{ik\cdot R}}{v_g(\boldsymbol{\mathrm{k}})},
\end{equation}
where $\ell_{\omega \left(\mathbf{k}\right)}$ is the isofrequency contour.

\subsubsection{2D free space}

\noindent We assume the dipole direction is normal to the 2D plane, the radiative decay rate is written as
\begin{equation}
\begin{split}
\Gamma_{\mathrm{Re}}\left(\omega \right)
&=\frac{\mu_1\mu_2\omega k}{4{\varepsilon }_0 c}\int^{\mathrm{2}\mathrm{\piup }}_0{d\theta e^{ikR{\mathrm{cos} \theta}}} \\
 &=B^{\left(2D\right)}\frac{k}{c}J_0\left(kR\right),
 \end{split}
\end{equation}
where $J_0\left(x\right)$ is the zero-order of the Bessel functions of the first kind and $B^{\left(2D\right)}={{\mu}_1{\mu}_2\omega}/{4{\varepsilon }_0 }$ is a frequency-related parameter.

\subsubsection{2D linear dispersion near Dirac point}

The dispersion relation near a Dirac point is written as
\begin{equation}
\omega_{\boldsymbol{\mathrm{k}}}=v\left|\boldsymbol{\mathrm{k}}\boldsymbol{\mathrm{-}}{\mathbf{k}}_{\boldsymbol{\mathrm{c}}}\right|+{\omega }_c=vq+{\omega }_c.
\end{equation}
Then the group velocity is written as $v_g=v$. The radiative interaction is given by
\begin{equation}
\begin{split}
{\mathrm{\Gamma }}_{\mathrm{Re}}\left(\omega \right)
&=\frac{{\mu }_1{\mu }_2\omega q}{4{\varepsilon }_0v}e^{i{\boldsymbol{\mathrm{k}}}_c\cdot \boldsymbol{\mathrm{R}}}\int^{\mathrm{2}\mathrm{\piup }}_0{d{\theta }_qe^{iqR{\mathrm{cos} {\theta }_q\ }}} \\
&=B^{\left(2D\right)}\frac{q}{v}J_0\left(qR\right)e^{i{\boldsymbol{\mathrm{k}}}_c\cdot \boldsymbol{\mathrm{R}}}.
\end{split}
\end{equation}
The 2D Dirac points provide stronger and longer-range interaction than Weyl points because of the reduced dimensionality.

\subsubsection{2D quadratic dispersion near band edges.}

If the 2D dispersion relation is quadratic such as
\begin{equation}
{\omega}_{\mathbf{k}}=\beta {\left|\boldsymbol{\mathrm{k}}\boldsymbol{\mathrm{-}}{\boldsymbol{\mathrm{k}}}_{\boldsymbol{\mathrm{c}}}\right|}^2+{\omega }_c=\beta q^2+{\omega }_c
\end{equation}
and the group velocity is $v_g=2\beta q$. Then,
\begin{equation}
\begin{split}
\Gamma_{\mathrm{Re}}\left(\omega \right)
& = \frac{{\mu}_1{\mu }_2\omega}{8{\varepsilon }_0\hslash \beta }e^{i{\boldsymbol{\mathrm{k}}}_c\cdot \boldsymbol{\mathrm{R}}}\int^{\mathrm{2}\mathrm{\piup }}_0{d{\theta }_qe^{iqR{\mathrm{cos} {\theta }_q\ }}} \\
&=B^{\left(2D\right)}\frac{1}{2\beta }J_0\left(qR\right)e^{i{\boldsymbol{\mathrm{k}}}_c\cdot \boldsymbol{\mathrm{R}}}.
\end{split}
\end{equation}
In this case, the strength do not be limited by small $q$.
}

\section{The Weyl Photonic Crystal}\label{sec:Weyl_PC}


\subsection{Numerical method and details}
The index of polarization $\eta$ is replaced by the index of band $n$. If the transition frequency of TLSs $\omega_0$ is in the $n_0$th band, the cooperative decay rate is given by
\begin{equation}
\begin{split}
\Gamma_\mathrm{Re} \left( \omega_0 \right)
= &
\frac{\omega_0}{16\pi^2\varepsilon_0}
\iint_{\mathcal{S}_{\omega_{0}(\mathbf{k})}}
 d\mathcal{S}_\mathbf{k}  \\
&\times
\left( \boldsymbol{\mathrm{\muup}}_1\cdot \boldsymbol{\mathrm{\epsilonup}}_{{n_0},\mathbf{k}} \right)^\ast
\left(
\boldsymbol{\mathrm{\muup}}_2 \cdot \boldsymbol{\mathrm{\epsilonup}}_{{n_0},\mathbf{k}} \right)
\frac{e^{i\mathbf{k} \cdot \mathbf{R} } }{ v_g\left(\mathbf{k},n_0 \right) }.
\end{split}
\end{equation}
The cooperative energy shift is written as
\begin{equation}\label{eq:shift_numeric}
\begin{split}
\Gamma_\mathrm{Im} (\omega_0)
=&
\frac{1}{16{\pi }^3 \varepsilon_0}
\sum_n
\mathbb{P}
\int_{\omega_n^\mathrm{min}}^{\omega_n^\mathrm{max}}  d\omega_n
\iint_{\mathcal{S}_{\omega_n(\mathbf{k})}} dS_\mathbf{k}
 \frac{\omega_{n}}{v_{n,\mathbf{k}} }  \\
 & \ \ \ \ \times
\Bigg(
 \left({\boldsymbol{\mathrm{\muup}}}_1\cdot {\boldsymbol{\mathrm{\epsilonup}}}_{n,\mathbf{k} }\right)^\ast
 {\left({\boldsymbol{\mathrm{\muup}}}_2\cdot {\boldsymbol{\mathrm{\epsilonup }}}_{n,\mathbf{k} }\right)}
\frac{e^{i\boldsymbol{\mathrm{k}}\cdot \mathbf{R} } }{\omega_{n} -\omega_0} \\
&  \ \ \ \  \ \ \ +
  \left({\boldsymbol{\mathrm{\muup}}}_1\cdot {\boldsymbol{\mathrm{\epsilonup}}}_{n,\mathbf{k} }\right) {\left({\boldsymbol{\mathrm{\muup}}}_2\cdot {\boldsymbol{\mathrm{\epsilonup }}}_{n,\mathbf{k} }\right)^\ast}
 \frac{e^{-i\boldsymbol{\mathrm{k}}\cdot \mathbf{R} } }{\omega_{n}+\omega_0}
\Bigg) \\
 \\
 =&
\frac{1}{\pi }
\sum_n
\mathbb{P}
\int_{\omega_n^\mathrm{min}}^{\omega_n^\mathrm{max}}  d\omega_n
 \\
& \ \ \ \  \times
\left( \frac{1 }{\omega_{n} -\omega_0}\Gamma_\mathrm{Re}(\omega_n)
 + \frac{ 1 }{\omega_{n}+\omega_0}\Gamma_\mathrm{Re}^\ast(\omega_n)
 \right).
\end{split}
\end{equation}

Numerically, we use the MPB software package~\cite{johnson2001block} to calculate eigen-modes of the Weyl photonic crystal in Fig. 3 of the main text. We set the resolution in the unit cell as $30\times30\times30$. Then, the frequency of the Weyl points is $\omega_\mathrm{wp}=0.55096 [2\pi c/a]$, which falls in between $4$th and $5$th bands.


\subsection{Dipole orientations and spatial placements of TLSs}

The details of the gyroid photonic crystal are shown in Fig. S1.
To demonstrate the effect of TLS dipole orientations, we fix the direction of the distance vector as $\hat{\mathbf{R}}=\left( -1,\ 1,\ 1 \right)/\sqrt{3}$. Fig.~\ref{fig:dipole} shows the real part of radiative interaction $\Gamma_\mathrm{Re}$ as a function of distance $R$ with arbitrary dipole orientations. Although the the dipole orientations affect the oscillation patterns of $\Gamma_\mathrm{Re}$ curves, the envelopes of all curves show negligible decay even at 30 wavelengths. In addition, the amplitudes of $\Gamma_\mathrm{Re}$ with different dipole orientations stay on a same order of magnitudes.

To show the effect of distance vector, the dipole orientations are fixed as $\hat{\boldsymbol{\mathrm{\muup}}}_{1,2}=\left( 0,\ 1,\ 0 \right)$. Fig.~\ref{fig:distance} shows the real part of radiative interaction $\Gamma_\mathrm{Re}$ as a function of distance $R$ with different distance vectors. Similarly to the effect of dipole orientations, the variation of $\hat{\mathbf{R}}$ only influences the oscillation patterns of $\Gamma_\mathrm{Re}$ curves. However, the envelopes of all curves exhibit negligible decay at 30 wavelengths. Similary, the variation of the first TLS location $\mathbf{r}_1$ in a unit cell does not affect the interaction range.

\section{ Interactions in other photonic environments }

\begin{figure}
\begin{center}
\epsfig{figure=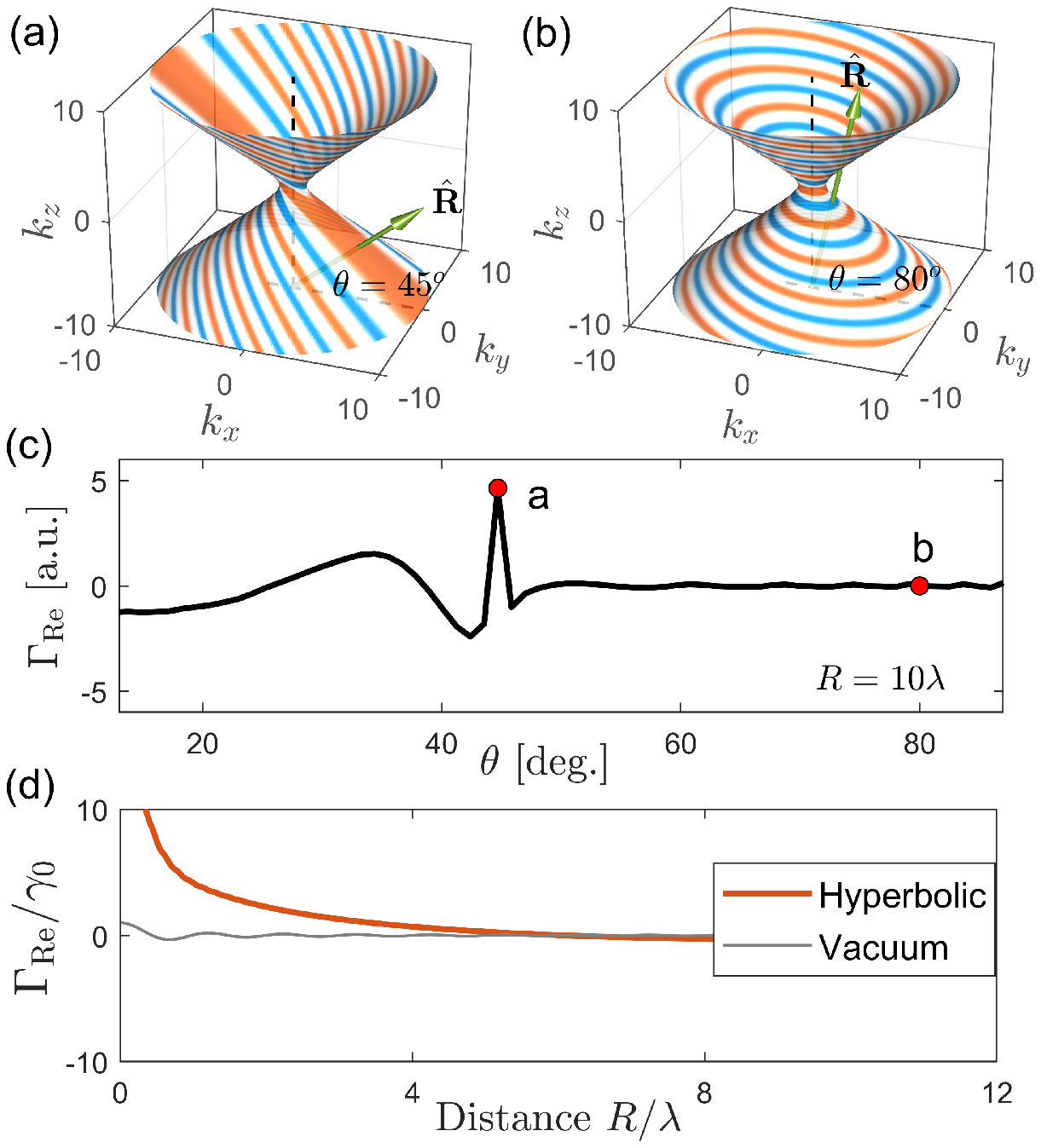,width=\linewidth}
\end{center}
\caption{
(a) \& (b) The real part of integrand $\rho_k e^{i\mathbf{k}\cdot \mathbf{R}}$ on the hyperbolic isosurface with $\theta=45^\circ$ and $80^\circ$, respectively.
The dipole orientations are $\hat{\boldsymbol{\mathrm{\muup}}}_{1,2} =\left(0,\ 1,\ 0 \right)$.
(c) $\Gamma_\mathrm{Re}$ as a function of $\theta$ with a fixed distance $R=10\lambda$. Red dots correspond to the cases in (a) and (b).
{(d) $\Gamma_\mathrm{Re}$ as a function of distance for the hyperbolic (red) and vacuum (gray) cases. $\gamma_0$ is the spontaneous decay rate in free space. The angle is fixed as $\theta=45^\circ$.}
}
\label{fig:hyperbolic}
\end{figure}

Our theory is also applicable for understanding the dipole-dipole interactions in other photonic environments, such as near the bandedge of photonic crystals, index-near-zero materials, hyperbolic materials, etc. Here, we describe the interaction in hyperbolic media as an especially interesting case. Its isosurface is infinitely large, which normally leads to a very short interaction range. This is indeed the case. However, because of its unique shape, for the direction $\mathbf{R}$ that is normal to the isosurface, the interaction range can be very long. This can also be clearly seen in the illustration of the integrand on the surface as we will show now. The dispersion relation of hyperbolic materials can be described by
\begin{equation}
\frac{k_x^2+k_y^2}{\varepsilon_z} + \frac{k_z^2}{\varepsilon_x} = \frac{\omega^2}{c^2}.
\end{equation}
Here, we choose the second type hyperbolic material with $\varepsilon_z=-\varepsilon_x=1$. The isosurface and the real part of integrand $\rho_\mathbf{k} e^{i\mathbf{k} \cdot \mathbf{R}}$ are shown in Fig.~\ref{fig:hyperbolic} (a) and (b) with $\theta=45^\circ$ and $80^\circ$, respectively. Here, $\theta$ is the included angle between $\hat{\mathbf{R}}$ and $x$-axis.
When $\hat{\mathbf{R}}$ is normal to the isosurface, i.e. $\theta=45^\circ$,there is a thick red (positive) strip in the oscillating pattern and it results in a large integral value.
If $\hat{\mathbf{R}}$ is aligned with a different direction, say $\theta=80^\circ$, the fast oscillation of $e^{i\mathbf{k}\cdot\mathbf{R}}$ results in cancellation of the integral and thus a weak interaction strength.
In Fig.~\ref{fig:hyperbolic}(c), we plot the cooperative decay rate at $R=10\lambda$ as a function of $\theta$. The left- and right-hand red dots correspond to the case in Fig.~\ref{fig:hyperbolic} (a) and (b), respectively. Our results agree with the exact numerical result in Ref.~\cite{cortes2017super}.

Furthermore, we also plot the radiative interaction $\Gamma_\mathrm{Re}$ as a function of distance for the hyperbolic material compared to the case in free space, as shown in Fig.~\ref{fig:hyperbolic} (d). It greatly agrees with the result in Ref.~\cite{cortes2017super}.


\begin{thebibliography}{37}%
\makeatletter
\providecommand \@ifxundefined [1]{%
 \@ifx{#1\undefined}
}%
\providecommand \@ifnum [1]{%
 \ifnum #1\expandafter \@firstoftwo
 \else \expandafter \@secondoftwo
 \fi
}%
\providecommand \@ifx [1]{%
 \ifx #1\expandafter \@firstoftwo
 \else \expandafter \@secondoftwo
 \fi
}%
\providecommand \natexlab [1]{#1}%
\providecommand \enquote  [1]{``#1''}%
\providecommand \bibnamefont  [1]{#1}%
\providecommand \bibfnamefont [1]{#1}%
\providecommand \citenamefont [1]{#1}%
\providecommand \href@noop [0]{\@secondoftwo}%
\providecommand \href [0]{\begingroup \@sanitize@url \@href}%
\providecommand \@href[1]{\@@startlink{#1}\@@href}%
\providecommand \@@href[1]{\endgroup#1\@@endlink}%
\providecommand \@sanitize@url [0]{\catcode `\\12\catcode `\$12\catcode
  `\&12\catcode `\#12\catcode `\^12\catcode `\_12\catcode `\%12\relax}%
\providecommand \@@startlink[1]{}%
\providecommand \@@endlink[0]{}%
\providecommand \url  [0]{\begingroup\@sanitize@url \@url }%
\providecommand \@url [1]{\endgroup\@href {#1}{\urlprefix }}%
\providecommand \urlprefix  [0]{URL }%
\providecommand \Eprint [0]{\href }%
\providecommand \doibase [0]{http://dx.doi.org/}%
\providecommand \selectlanguage [0]{\@gobble}%
\providecommand \bibinfo  [0]{\@secondoftwo}%
\providecommand \bibfield  [0]{\@secondoftwo}%
\providecommand \translation [1]{[#1]}%
\providecommand \BibitemOpen [0]{}%
\providecommand \bibitemStop [0]{}%
\providecommand \bibitemNoStop [0]{.\EOS\space}%
\providecommand \EOS [0]{\spacefactor3000\relax}%
\providecommand \BibitemShut  [1]{\csname bibitem#1\endcsname}%
\let\auto@bib@innerbib\@empty
\bibitem [{\citenamefont {Scully}\ and\ \citenamefont
  {Svidzinsky}(2009)}]{scully2009super}%
  \BibitemOpen
  \bibfield  {author} {\bibinfo {author} {\bibfnamefont {M.~O.}\ \bibnamefont
  {Scully}}\ and\ \bibinfo {author} {\bibfnamefont {A.~A.}\ \bibnamefont
  {Svidzinsky}},\ }\href@noop {} {\bibfield  {journal} {\bibinfo  {journal}
  {Science}\ }\textbf {\bibinfo {volume} {325}},\ \bibinfo {pages} {1510}
  (\bibinfo {year} {2009})}\BibitemShut {NoStop}%
\bibitem [{\citenamefont {Solano}\ \emph {et~al.}(2017)\citenamefont {Solano},
  \citenamefont {Barberis-Blostein}, \citenamefont {Fatemi}, \citenamefont
  {Orozco},\ and\ \citenamefont {Rolston}}]{solano2017super}%
  \BibitemOpen
  \bibfield  {author} {\bibinfo {author} {\bibfnamefont {P.}~\bibnamefont
  {Solano}}, \bibinfo {author} {\bibfnamefont {P.}~\bibnamefont
  {Barberis-Blostein}}, \bibinfo {author} {\bibfnamefont {F.~K.}\ \bibnamefont
  {Fatemi}}, \bibinfo {author} {\bibfnamefont {L.~A.}\ \bibnamefont {Orozco}},
  \ and\ \bibinfo {author} {\bibfnamefont {S.~L.}\ \bibnamefont {Rolston}},\
  }\href@noop {} {\bibfield  {journal} {\bibinfo  {journal} {Nature
  communications}\ }\textbf {\bibinfo {volume} {8}},\ \bibinfo {pages} {1857}
  (\bibinfo {year} {2017})}\BibitemShut {NoStop}%
\bibitem [{\citenamefont {Meir}\ \emph {et~al.}(2014)\citenamefont {Meir},
  \citenamefont {Schwartz}, \citenamefont {Shahmoon}, \citenamefont {Oron},\
  and\ \citenamefont {Ozeri}}]{meir2014cooperative}%
  \BibitemOpen
  \bibfield  {author} {\bibinfo {author} {\bibfnamefont {Z.}~\bibnamefont
  {Meir}}, \bibinfo {author} {\bibfnamefont {O.}~\bibnamefont {Schwartz}},
  \bibinfo {author} {\bibfnamefont {E.}~\bibnamefont {Shahmoon}}, \bibinfo
  {author} {\bibfnamefont {D.}~\bibnamefont {Oron}}, \ and\ \bibinfo {author}
  {\bibfnamefont {R.}~\bibnamefont {Ozeri}},\ }\href {\doibase
  10.1103/PhysRevLett.113.193002} {\bibfield  {journal} {\bibinfo  {journal}
  {Phys. Rev. Lett.}\ }\textbf {\bibinfo {volume} {113}},\ \bibinfo {pages}
  {193002} (\bibinfo {year} {2014})}\BibitemShut {NoStop}%
\bibitem [{\citenamefont {Clegg}(1995)}]{clegg1995fluorescence}%
  \BibitemOpen
  \bibfield  {author} {\bibinfo {author} {\bibfnamefont {R.~M.}\ \bibnamefont
  {Clegg}},\ }\href@noop {} {\bibfield  {journal} {\bibinfo  {journal} {Current
  opinion in biotechnology}\ }\textbf {\bibinfo {volume} {6}},\ \bibinfo
  {pages} {103} (\bibinfo {year} {1995})}\BibitemShut {NoStop}%
\bibitem [{\citenamefont {Garcia-Vidal}\ and\ \citenamefont
  {Feist}(2017)}]{garcia2017long}%
  \BibitemOpen
  \bibfield  {author} {\bibinfo {author} {\bibfnamefont {F.~J.}\ \bibnamefont
  {Garcia-Vidal}}\ and\ \bibinfo {author} {\bibfnamefont {J.}~\bibnamefont
  {Feist}},\ }\href@noop {} {\bibfield  {journal} {\bibinfo  {journal}
  {Science}\ }\textbf {\bibinfo {volume} {357}},\ \bibinfo {pages} {1357}
  (\bibinfo {year} {2017})}\BibitemShut {NoStop}%
\bibitem [{\citenamefont {Van~Loo}\ \emph {et~al.}(2013)\citenamefont
  {Van~Loo}, \citenamefont {Fedorov}, \citenamefont {Lalumi{\`e}re},
  \citenamefont {Sanders}, \citenamefont {Blais},\ and\ \citenamefont
  {Wallraff}}]{van2013photon}%
  \BibitemOpen
  \bibfield  {author} {\bibinfo {author} {\bibfnamefont {A.~F.}\ \bibnamefont
  {Van~Loo}}, \bibinfo {author} {\bibfnamefont {A.}~\bibnamefont {Fedorov}},
  \bibinfo {author} {\bibfnamefont {K.}~\bibnamefont {Lalumi{\`e}re}}, \bibinfo
  {author} {\bibfnamefont {B.~C.}\ \bibnamefont {Sanders}}, \bibinfo {author}
  {\bibfnamefont {A.}~\bibnamefont {Blais}}, \ and\ \bibinfo {author}
  {\bibfnamefont {A.}~\bibnamefont {Wallraff}},\ }\href@noop {} {\bibfield
  {journal} {\bibinfo  {journal} {Science}\ }\textbf {\bibinfo {volume}
  {342}},\ \bibinfo {pages} {1494} (\bibinfo {year} {2013})}\BibitemShut
  {NoStop}%
\bibitem [{\citenamefont {Burkard}\ and\ \citenamefont
  {Imamoglu}(2006)}]{burkard2006ultra}%
  \BibitemOpen
  \bibfield  {author} {\bibinfo {author} {\bibfnamefont {G.}~\bibnamefont
  {Burkard}}\ and\ \bibinfo {author} {\bibfnamefont {A.}~\bibnamefont
  {Imamoglu}},\ }\href@noop {} {\bibfield  {journal} {\bibinfo  {journal}
  {Physical Review B}\ }\textbf {\bibinfo {volume} {74}},\ \bibinfo {pages}
  {041307} (\bibinfo {year} {2006})}\BibitemShut {NoStop}%
\bibitem [{\citenamefont {Petrosyan}\ and\ \citenamefont
  {Fleischhauer}(2008)}]{petrosyan2008quantum}%
  \BibitemOpen
  \bibfield  {author} {\bibinfo {author} {\bibfnamefont {D.}~\bibnamefont
  {Petrosyan}}\ and\ \bibinfo {author} {\bibfnamefont {M.}~\bibnamefont
  {Fleischhauer}},\ }\href@noop {} {\bibfield  {journal} {\bibinfo  {journal}
  {Physical review letters}\ }\textbf {\bibinfo {volume} {100}},\ \bibinfo
  {pages} {170501} (\bibinfo {year} {2008})}\BibitemShut {NoStop}%
\bibitem [{\citenamefont {Mingaleev}\ \emph {et~al.}(2000)\citenamefont
  {Mingaleev}, \citenamefont {Kivshar},\ and\ \citenamefont
  {Sammut}}]{mingaleev2000long}%
  \BibitemOpen
  \bibfield  {author} {\bibinfo {author} {\bibfnamefont {S.~F.}\ \bibnamefont
  {Mingaleev}}, \bibinfo {author} {\bibfnamefont {Y.~S.}\ \bibnamefont
  {Kivshar}}, \ and\ \bibinfo {author} {\bibfnamefont {R.~A.}\ \bibnamefont
  {Sammut}},\ }\href@noop {} {\bibfield  {journal} {\bibinfo  {journal}
  {Physical Review E}\ }\textbf {\bibinfo {volume} {62}},\ \bibinfo {pages}
  {5777} (\bibinfo {year} {2000})}\BibitemShut {NoStop}%
\bibitem [{\citenamefont {Gonzalez-Tudela}\ \emph {et~al.}(2011)\citenamefont
  {Gonzalez-Tudela}, \citenamefont {Martin-Cano}, \citenamefont {Moreno},
  \citenamefont {Martin-Moreno}, \citenamefont {Tejedor},\ and\ \citenamefont
  {Garcia-Vidal}}]{gonzalez2011entanglement}%
  \BibitemOpen
  \bibfield  {author} {\bibinfo {author} {\bibfnamefont {A.}~\bibnamefont
  {Gonzalez-Tudela}}, \bibinfo {author} {\bibfnamefont {D.}~\bibnamefont
  {Martin-Cano}}, \bibinfo {author} {\bibfnamefont {E.}~\bibnamefont {Moreno}},
  \bibinfo {author} {\bibfnamefont {L.}~\bibnamefont {Martin-Moreno}}, \bibinfo
  {author} {\bibfnamefont {C.}~\bibnamefont {Tejedor}}, \ and\ \bibinfo
  {author} {\bibfnamefont {F.~J.}\ \bibnamefont {Garcia-Vidal}},\ }\href@noop
  {} {\bibfield  {journal} {\bibinfo  {journal} {Physical review letters}\
  }\textbf {\bibinfo {volume} {106}},\ \bibinfo {pages} {020501} (\bibinfo
  {year} {2011})}\BibitemShut {NoStop}%
\bibitem [{\citenamefont {Shahmoon}\ and\ \citenamefont
  {Kurizki}(2013)}]{shahmoon2013nonradiative}%
  \BibitemOpen
  \bibfield  {author} {\bibinfo {author} {\bibfnamefont {E.}~\bibnamefont
  {Shahmoon}}\ and\ \bibinfo {author} {\bibfnamefont {G.}~\bibnamefont
  {Kurizki}},\ }\href@noop {} {\bibfield  {journal} {\bibinfo  {journal}
  {Physical Review A}\ }\textbf {\bibinfo {volume} {87}},\ \bibinfo {pages}
  {033831} (\bibinfo {year} {2013})}\BibitemShut {NoStop}%
\bibitem [{\citenamefont {Hood}\ \emph {et~al.}(2016)\citenamefont {Hood},
  \citenamefont {Goban}, \citenamefont {Asenjo-Garcia}, \citenamefont {Lu},
  \citenamefont {Yu}, \citenamefont {Chang},\ and\ \citenamefont
  {Kimble}}]{hood2016atom}%
  \BibitemOpen
  \bibfield  {author} {\bibinfo {author} {\bibfnamefont {J.~D.}\ \bibnamefont
  {Hood}}, \bibinfo {author} {\bibfnamefont {A.}~\bibnamefont {Goban}},
  \bibinfo {author} {\bibfnamefont {A.}~\bibnamefont {Asenjo-Garcia}}, \bibinfo
  {author} {\bibfnamefont {M.}~\bibnamefont {Lu}}, \bibinfo {author}
  {\bibfnamefont {S.-P.}\ \bibnamefont {Yu}}, \bibinfo {author} {\bibfnamefont
  {D.~E.}\ \bibnamefont {Chang}}, \ and\ \bibinfo {author} {\bibfnamefont
  {H.}~\bibnamefont {Kimble}},\ }\href@noop {} {\bibfield  {journal} {\bibinfo
  {journal} {Proceedings of the National Academy of Sciences}\ }\textbf
  {\bibinfo {volume} {113}},\ \bibinfo {pages} {10507} (\bibinfo {year}
  {2016})}\BibitemShut {NoStop}%
\bibitem [{\citenamefont {Imamog}\ \emph {et~al.}(1999)\citenamefont {Imamog},
  \citenamefont {Awschalom}, \citenamefont {Burkard}, \citenamefont
  {DiVincenzo}, \citenamefont {Loss}, \citenamefont {Sherwin}, \citenamefont
  {Small} \emph {et~al.}}]{imamog1999quantum}%
  \BibitemOpen
  \bibfield  {author} {\bibinfo {author} {\bibfnamefont {A.}~\bibnamefont
  {Imamog}}, \bibinfo {author} {\bibfnamefont {D.~D.}\ \bibnamefont
  {Awschalom}}, \bibinfo {author} {\bibfnamefont {G.}~\bibnamefont {Burkard}},
  \bibinfo {author} {\bibfnamefont {D.~P.}\ \bibnamefont {DiVincenzo}},
  \bibinfo {author} {\bibfnamefont {D.}~\bibnamefont {Loss}}, \bibinfo {author}
  {\bibfnamefont {M.}~\bibnamefont {Sherwin}}, \bibinfo {author} {\bibfnamefont
  {A.}~\bibnamefont {Small}},  \emph {et~al.},\ }\href@noop {} {\bibfield
  {journal} {\bibinfo  {journal} {Physical review letters}\ }\textbf {\bibinfo
  {volume} {83}},\ \bibinfo {pages} {4204} (\bibinfo {year}
  {1999})}\BibitemShut {NoStop}%
\bibitem [{\citenamefont {Maxwell}\ \emph {et~al.}(2013)\citenamefont
  {Maxwell}, \citenamefont {Szwer}, \citenamefont {Paredes-Barato},
  \citenamefont {Busche}, \citenamefont {Pritchard}, \citenamefont {Gauguet},
  \citenamefont {Weatherill}, \citenamefont {Jones},\ and\ \citenamefont
  {Adams}}]{maxwell2013storage}%
  \BibitemOpen
  \bibfield  {author} {\bibinfo {author} {\bibfnamefont {D.}~\bibnamefont
  {Maxwell}}, \bibinfo {author} {\bibfnamefont {D.}~\bibnamefont {Szwer}},
  \bibinfo {author} {\bibfnamefont {D.}~\bibnamefont {Paredes-Barato}},
  \bibinfo {author} {\bibfnamefont {H.}~\bibnamefont {Busche}}, \bibinfo
  {author} {\bibfnamefont {J.~D.}\ \bibnamefont {Pritchard}}, \bibinfo {author}
  {\bibfnamefont {A.}~\bibnamefont {Gauguet}}, \bibinfo {author} {\bibfnamefont
  {K.~J.}\ \bibnamefont {Weatherill}}, \bibinfo {author} {\bibfnamefont
  {M.}~\bibnamefont {Jones}}, \ and\ \bibinfo {author} {\bibfnamefont {C.~S.}\
  \bibnamefont {Adams}},\ }\href@noop {} {\bibfield  {journal} {\bibinfo
  {journal} {Physical review letters}\ }\textbf {\bibinfo {volume} {110}},\
  \bibinfo {pages} {103001} (\bibinfo {year} {2013})}\BibitemShut {NoStop}%
\bibitem [{\citenamefont {Kurizki}(1990)}]{kurizki1990two}%
  \BibitemOpen
  \bibfield  {author} {\bibinfo {author} {\bibfnamefont {G.}~\bibnamefont
  {Kurizki}},\ }\href@noop {} {\bibfield  {journal} {\bibinfo  {journal}
  {Physical Review A}\ }\textbf {\bibinfo {volume} {42}},\ \bibinfo {pages}
  {2915} (\bibinfo {year} {1990})}\BibitemShut {NoStop}%
\bibitem [{\citenamefont {Douglas}\ \emph {et~al.}(2015)\citenamefont
  {Douglas}, \citenamefont {Habibian}, \citenamefont {Hung}, \citenamefont
  {Gorshkov}, \citenamefont {Kimble},\ and\ \citenamefont
  {Chang}}]{douglas2015quantum}%
  \BibitemOpen
  \bibfield  {author} {\bibinfo {author} {\bibfnamefont {J.~S.}\ \bibnamefont
  {Douglas}}, \bibinfo {author} {\bibfnamefont {H.}~\bibnamefont {Habibian}},
  \bibinfo {author} {\bibfnamefont {C.-L.}\ \bibnamefont {Hung}}, \bibinfo
  {author} {\bibfnamefont {A.~V.}\ \bibnamefont {Gorshkov}}, \bibinfo {author}
  {\bibfnamefont {H.~J.}\ \bibnamefont {Kimble}}, \ and\ \bibinfo {author}
  {\bibfnamefont {D.~E.}\ \bibnamefont {Chang}},\ }\href@noop {} {\bibfield
  {journal} {\bibinfo  {journal} {Nature Photonics}\ }\textbf {\bibinfo
  {volume} {9}},\ \bibinfo {pages} {326} (\bibinfo {year} {2015})}\BibitemShut
  {NoStop}%
\bibitem [{\citenamefont {Notararigo}\ \emph {et~al.}(2018)\citenamefont
  {Notararigo}, \citenamefont {Passante},\ and\ \citenamefont
  {Rizzuto}}]{notararigo2018resonance}%
  \BibitemOpen
  \bibfield  {author} {\bibinfo {author} {\bibfnamefont {V.}~\bibnamefont
  {Notararigo}}, \bibinfo {author} {\bibfnamefont {R.}~\bibnamefont
  {Passante}}, \ and\ \bibinfo {author} {\bibfnamefont {L.}~\bibnamefont
  {Rizzuto}},\ }\href@noop {} {\bibfield  {journal} {\bibinfo  {journal}
  {Scientific reports}\ }\textbf {\bibinfo {volume} {8}},\ \bibinfo {pages}
  {5193} (\bibinfo {year} {2018})}\BibitemShut {NoStop}%
\bibitem [{\citenamefont {Fleury}\ and\ \citenamefont
  {Alu}(2013)}]{fleury2013enhanced}%
  \BibitemOpen
  \bibfield  {author} {\bibinfo {author} {\bibfnamefont {R.}~\bibnamefont
  {Fleury}}\ and\ \bibinfo {author} {\bibfnamefont {A.}~\bibnamefont {Alu}},\
  }\href@noop {} {\bibfield  {journal} {\bibinfo  {journal} {Physical Review
  B}\ }\textbf {\bibinfo {volume} {87}},\ \bibinfo {pages} {201101} (\bibinfo
  {year} {2013})}\BibitemShut {NoStop}%
\bibitem [{\citenamefont {Mahmoud}\ \emph {et~al.}(2017)\citenamefont
  {Mahmoud}, \citenamefont {Liberal},\ and\ \citenamefont
  {Engheta}}]{mahmoud2017dipole}%
  \BibitemOpen
  \bibfield  {author} {\bibinfo {author} {\bibfnamefont {A.}~\bibnamefont
  {Mahmoud}}, \bibinfo {author} {\bibfnamefont {I.}~\bibnamefont {Liberal}}, \
  and\ \bibinfo {author} {\bibfnamefont {N.}~\bibnamefont {Engheta}},\
  }\href@noop {} {\bibfield  {journal} {\bibinfo  {journal} {Optical Materials
  Express}\ }\textbf {\bibinfo {volume} {7}},\ \bibinfo {pages} {415} (\bibinfo
  {year} {2017})}\BibitemShut {NoStop}%
\bibitem [{\citenamefont {Liberal}\ and\ \citenamefont
  {Engheta}(2018)}]{liberal2018multiqubit}%
  \BibitemOpen
  \bibfield  {author} {\bibinfo {author} {\bibfnamefont {I.}~\bibnamefont
  {Liberal}}\ and\ \bibinfo {author} {\bibfnamefont {N.}~\bibnamefont
  {Engheta}},\ }\href@noop {} {\bibfield  {journal} {\bibinfo  {journal}
  {Physical Review A}\ }\textbf {\bibinfo {volume} {97}},\ \bibinfo {pages}
  {022309} (\bibinfo {year} {2018})}\BibitemShut {NoStop}%
\bibitem [{\citenamefont {Gundogdu}\ \emph {et~al.}(2015)\citenamefont
  {Gundogdu}, \citenamefont {Serebryannikov}, \citenamefont {Cakmak},\ and\
  \citenamefont {Ozbay}}]{gundogdu2015asymmetric}%
  \BibitemOpen
  \bibfield  {author} {\bibinfo {author} {\bibfnamefont {F.~T.}\ \bibnamefont
  {Gundogdu}}, \bibinfo {author} {\bibfnamefont {A.~E.}\ \bibnamefont
  {Serebryannikov}}, \bibinfo {author} {\bibfnamefont {A.~O.}\ \bibnamefont
  {Cakmak}}, \ and\ \bibinfo {author} {\bibfnamefont {E.}~\bibnamefont
  {Ozbay}},\ }\href@noop {} {\bibfield  {journal} {\bibinfo  {journal} {Optics
  express}\ }\textbf {\bibinfo {volume} {23}},\ \bibinfo {pages} {24120}
  (\bibinfo {year} {2015})}\BibitemShut {NoStop}%
\bibitem [{\citenamefont {Serebryannikov}\ \emph {et~al.}(2019)\citenamefont
  {Serebryannikov}, \citenamefont {Hajian}, \citenamefont {Krawczyk},
  \citenamefont {Vandenbosch},\ and\ \citenamefont
  {Ozbay}}]{serebryannikov2019embedded}%
  \BibitemOpen
  \bibfield  {author} {\bibinfo {author} {\bibfnamefont {A.~E.}\ \bibnamefont
  {Serebryannikov}}, \bibinfo {author} {\bibfnamefont {H.}~\bibnamefont
  {Hajian}}, \bibinfo {author} {\bibfnamefont {M.}~\bibnamefont {Krawczyk}},
  \bibinfo {author} {\bibfnamefont {G.~A.}\ \bibnamefont {Vandenbosch}}, \ and\
  \bibinfo {author} {\bibfnamefont {E.}~\bibnamefont {Ozbay}},\ }\href@noop {}
  {\bibfield  {journal} {\bibinfo  {journal} {Optical Materials Express}\
  }\textbf {\bibinfo {volume} {9}},\ \bibinfo {pages} {3169} (\bibinfo {year}
  {2019})}\BibitemShut {NoStop}%
\bibitem [{\citenamefont {Sato}\ \emph {et~al.}(2012)\citenamefont {Sato},
  \citenamefont {Tanaka}, \citenamefont {Upham}, \citenamefont {Takahashi},
  \citenamefont {Asano},\ and\ \citenamefont {Noda}}]{sato2012strong}%
  \BibitemOpen
  \bibfield  {author} {\bibinfo {author} {\bibfnamefont {Y.}~\bibnamefont
  {Sato}}, \bibinfo {author} {\bibfnamefont {Y.}~\bibnamefont {Tanaka}},
  \bibinfo {author} {\bibfnamefont {J.}~\bibnamefont {Upham}}, \bibinfo
  {author} {\bibfnamefont {Y.}~\bibnamefont {Takahashi}}, \bibinfo {author}
  {\bibfnamefont {T.}~\bibnamefont {Asano}}, \ and\ \bibinfo {author}
  {\bibfnamefont {S.}~\bibnamefont {Noda}},\ }\href@noop {} {\bibfield
  {journal} {\bibinfo  {journal} {Nature Photonics}\ }\textbf {\bibinfo
  {volume} {6}},\ \bibinfo {pages} {56} (\bibinfo {year} {2012})}\BibitemShut
  {NoStop}%
\bibitem [{\citenamefont {Le~Kien}\ and\ \citenamefont
  {Rauschenbeutel}(2017)}]{le2017nanofiber}%
  \BibitemOpen
  \bibfield  {author} {\bibinfo {author} {\bibfnamefont {F.}~\bibnamefont
  {Le~Kien}}\ and\ \bibinfo {author} {\bibfnamefont {A.}~\bibnamefont
  {Rauschenbeutel}},\ }\href@noop {} {\bibfield  {journal} {\bibinfo  {journal}
  {Physical Review A}\ }\textbf {\bibinfo {volume} {95}},\ \bibinfo {pages}
  {023838} (\bibinfo {year} {2017})}\BibitemShut {NoStop}%
\bibitem [{\citenamefont {Lecamp}\ \emph {et~al.}(2007)\citenamefont {Lecamp},
  \citenamefont {Lalanne},\ and\ \citenamefont {Hugonin}}]{lecamp2007very}%
  \BibitemOpen
  \bibfield  {author} {\bibinfo {author} {\bibfnamefont {G.}~\bibnamefont
  {Lecamp}}, \bibinfo {author} {\bibfnamefont {P.}~\bibnamefont {Lalanne}}, \
  and\ \bibinfo {author} {\bibfnamefont {J.}~\bibnamefont {Hugonin}},\
  }\href@noop {} {\bibfield  {journal} {\bibinfo  {journal} {Physical review
  letters}\ }\textbf {\bibinfo {volume} {99}},\ \bibinfo {pages} {023902}
  (\bibinfo {year} {2007})}\BibitemShut {NoStop}%
\bibitem [{\citenamefont {Hughes}(2007)}]{hughes2007coupled}%
  \BibitemOpen
  \bibfield  {author} {\bibinfo {author} {\bibfnamefont {S.}~\bibnamefont
  {Hughes}},\ }\href@noop {} {\bibfield  {journal} {\bibinfo  {journal}
  {Physical review letters}\ }\textbf {\bibinfo {volume} {98}},\ \bibinfo
  {pages} {083603} (\bibinfo {year} {2007})}\BibitemShut {NoStop}%
\bibitem [{\citenamefont {Yao}\ and\ \citenamefont
  {Hughes}(2009)}]{yao2009macroscopic}%
  \BibitemOpen
  \bibfield  {author} {\bibinfo {author} {\bibfnamefont {P.}~\bibnamefont
  {Yao}}\ and\ \bibinfo {author} {\bibfnamefont {S.}~\bibnamefont {Hughes}},\
  }\href@noop {} {\bibfield  {journal} {\bibinfo  {journal} {Optics Express}\
  }\textbf {\bibinfo {volume} {17}},\ \bibinfo {pages} {11505} (\bibinfo {year}
  {2009})}\BibitemShut {NoStop}%
\bibitem [{\citenamefont {Minkov}\ and\ \citenamefont
  {Savona}(2013)}]{minkov2013radiative}%
  \BibitemOpen
  \bibfield  {author} {\bibinfo {author} {\bibfnamefont {M.}~\bibnamefont
  {Minkov}}\ and\ \bibinfo {author} {\bibfnamefont {V.}~\bibnamefont
  {Savona}},\ }\href@noop {} {\bibfield  {journal} {\bibinfo  {journal}
  {Physical Review B}\ }\textbf {\bibinfo {volume} {87}},\ \bibinfo {pages}
  {125306} (\bibinfo {year} {2013})}\BibitemShut {NoStop}%
\bibitem [{\citenamefont {Hung}\ \emph {et~al.}(2013)\citenamefont {Hung},
  \citenamefont {Meenehan}, \citenamefont {Chang}, \citenamefont {Painter},\
  and\ \citenamefont {Kimble}}]{hung2013trapped}%
  \BibitemOpen
  \bibfield  {author} {\bibinfo {author} {\bibfnamefont {C.}~\bibnamefont
  {Hung}}, \bibinfo {author} {\bibfnamefont {S.}~\bibnamefont {Meenehan}},
  \bibinfo {author} {\bibfnamefont {D.}~\bibnamefont {Chang}}, \bibinfo
  {author} {\bibfnamefont {O.}~\bibnamefont {Painter}}, \ and\ \bibinfo
  {author} {\bibfnamefont {H.}~\bibnamefont {Kimble}},\ }\href@noop {}
  {\bibfield  {journal} {\bibinfo  {journal} {New Journal of Physics}\ }\textbf
  {\bibinfo {volume} {15}},\ \bibinfo {pages} {083026} (\bibinfo {year}
  {2013})}\BibitemShut {NoStop}%
\bibitem [{\citenamefont {Vasco}\ \emph {et~al.}(2014)\citenamefont {Vasco},
  \citenamefont {Guimaraes},\ and\ \citenamefont {Gerace}}]{vasco2014long}%
  \BibitemOpen
  \bibfield  {author} {\bibinfo {author} {\bibfnamefont {J.}~\bibnamefont
  {Vasco}}, \bibinfo {author} {\bibfnamefont {P.}~\bibnamefont {Guimaraes}}, \
  and\ \bibinfo {author} {\bibfnamefont {D.}~\bibnamefont {Gerace}},\
  }\href@noop {} {\bibfield  {journal} {\bibinfo  {journal} {Physical Review
  B}\ }\textbf {\bibinfo {volume} {90}},\ \bibinfo {pages} {155436} (\bibinfo
  {year} {2014})}\BibitemShut {NoStop}%
\bibitem [{\citenamefont {Biehs}\ \emph {et~al.}(2016)\citenamefont {Biehs},
  \citenamefont {Menon},\ and\ \citenamefont {Agarwal}}]{biehs2016long}%
  \BibitemOpen
  \bibfield  {author} {\bibinfo {author} {\bibfnamefont {S.-A.}\ \bibnamefont
  {Biehs}}, \bibinfo {author} {\bibfnamefont {V.~M.}\ \bibnamefont {Menon}}, \
  and\ \bibinfo {author} {\bibfnamefont {G.}~\bibnamefont {Agarwal}},\
  }\href@noop {} {\bibfield  {journal} {\bibinfo  {journal} {Physical Review
  B}\ }\textbf {\bibinfo {volume} {93}},\ \bibinfo {pages} {245439} (\bibinfo
  {year} {2016})}\BibitemShut {NoStop}%
\bibitem [{\citenamefont {Cortes}\ and\ \citenamefont
  {Jacob}(2017)}]{cortes2017super}%
  \BibitemOpen
  \bibfield  {author} {\bibinfo {author} {\bibfnamefont {C.~L.}\ \bibnamefont
  {Cortes}}\ and\ \bibinfo {author} {\bibfnamefont {Z.}~\bibnamefont {Jacob}},\
  }\href@noop {} {\bibfield  {journal} {\bibinfo  {journal} {Nature
  communications}\ }\textbf {\bibinfo {volume} {8}},\ \bibinfo {pages} {14144}
  (\bibinfo {year} {2017})}\BibitemShut {NoStop}%
\bibitem [{\citenamefont {Bay}\ \emph {et~al.}(1997)\citenamefont {Bay},
  \citenamefont {Lambropoulos},\ and\ \citenamefont {M{\o}lmer}}]{bay1997atom}%
  \BibitemOpen
  \bibfield  {author} {\bibinfo {author} {\bibfnamefont {S.}~\bibnamefont
  {Bay}}, \bibinfo {author} {\bibfnamefont {P.}~\bibnamefont {Lambropoulos}}, \
  and\ \bibinfo {author} {\bibfnamefont {K.}~\bibnamefont {M{\o}lmer}},\
  }\href@noop {} {\bibfield  {journal} {\bibinfo  {journal} {Physical Review
  A}\ }\textbf {\bibinfo {volume} {55}},\ \bibinfo {pages} {1485} (\bibinfo
  {year} {1997})}\BibitemShut {NoStop}%
\bibitem [{\citenamefont {Yang}\ \emph {et~al.}(2018)\citenamefont {Yang},
  \citenamefont {Guo}, \citenamefont {Tremain}, \citenamefont {Liu},
  \citenamefont {Barr}, \citenamefont {Yan}, \citenamefont {Gao}, \citenamefont
  {Liu}, \citenamefont {Xiang}, \citenamefont {Chen} \emph
  {et~al.}}]{yang2018ideal}%
  \BibitemOpen
  \bibfield  {author} {\bibinfo {author} {\bibfnamefont {B.}~\bibnamefont
  {Yang}}, \bibinfo {author} {\bibfnamefont {Q.}~\bibnamefont {Guo}}, \bibinfo
  {author} {\bibfnamefont {B.}~\bibnamefont {Tremain}}, \bibinfo {author}
  {\bibfnamefont {R.}~\bibnamefont {Liu}}, \bibinfo {author} {\bibfnamefont
  {L.~E.}\ \bibnamefont {Barr}}, \bibinfo {author} {\bibfnamefont
  {Q.}~\bibnamefont {Yan}}, \bibinfo {author} {\bibfnamefont {W.}~\bibnamefont
  {Gao}}, \bibinfo {author} {\bibfnamefont {H.}~\bibnamefont {Liu}}, \bibinfo
  {author} {\bibfnamefont {Y.}~\bibnamefont {Xiang}}, \bibinfo {author}
  {\bibfnamefont {J.}~\bibnamefont {Chen}},  \emph {et~al.},\ }\href@noop {}
  {\bibfield  {journal} {\bibinfo  {journal} {Science}\ }\textbf {\bibinfo
  {volume} {359}},\ \bibinfo {pages} {1013} (\bibinfo {year}
  {2018})}\BibitemShut {NoStop}%
\bibitem [{\citenamefont {Lu}\ \emph {et~al.}(2013)\citenamefont {Lu},
  \citenamefont {Fu}, \citenamefont {Joannopoulos},\ and\ \citenamefont
  {Solja{\v{c}}i{\'c}}}]{lu2013weyl}%
  \BibitemOpen
  \bibfield  {author} {\bibinfo {author} {\bibfnamefont {L.}~\bibnamefont
  {Lu}}, \bibinfo {author} {\bibfnamefont {L.}~\bibnamefont {Fu}}, \bibinfo
  {author} {\bibfnamefont {J.~D.}\ \bibnamefont {Joannopoulos}}, \ and\
  \bibinfo {author} {\bibfnamefont {M.}~\bibnamefont {Solja{\v{c}}i{\'c}}},\
  }\href@noop {} {\bibfield  {journal} {\bibinfo  {journal} {Nature photonics}\
  }\textbf {\bibinfo {volume} {7}},\ \bibinfo {pages} {294} (\bibinfo {year}
  {2013})}\BibitemShut {NoStop}%
\bibitem [{\citenamefont {Wang}\ \emph {et~al.}(2016)\citenamefont {Wang},
  \citenamefont {Jian},\ and\ \citenamefont {Yao}}]{wang2016topological}%
  \BibitemOpen
  \bibfield  {author} {\bibinfo {author} {\bibfnamefont {L.}~\bibnamefont
  {Wang}}, \bibinfo {author} {\bibfnamefont {S.-K.}\ \bibnamefont {Jian}}, \
  and\ \bibinfo {author} {\bibfnamefont {H.}~\bibnamefont {Yao}},\ }\href@noop
  {} {\bibfield  {journal} {\bibinfo  {journal} {Physical Review A}\ }\textbf
  {\bibinfo {volume} {93}},\ \bibinfo {pages} {061801} (\bibinfo {year}
  {2016})}\BibitemShut {NoStop}%
\bibitem [{\citenamefont {Johnson}\ and\ \citenamefont
  {Joannopoulos}(2001)}]{johnson2001block}%
  \BibitemOpen
  \bibfield  {author} {\bibinfo {author} {\bibfnamefont {S.~G.}\ \bibnamefont
  {Johnson}}\ and\ \bibinfo {author} {\bibfnamefont {J.~D.}\ \bibnamefont
  {Joannopoulos}},\ }\href@noop {} {\bibfield  {journal} {\bibinfo  {journal}
  {Optics express}\ }\textbf {\bibinfo {volume} {8}},\ \bibinfo {pages} {173}
  (\bibinfo {year} {2001})}\BibitemShut {NoStop}%
\end{thebibliography}
\end{document}